\documentclass[
aps,
prl,
showpacs,
letterpaper,
twocolumn,
preprint
longbibliography,
superscriptaddress,
notitlepage,
nofootinbib,
floatfix
]{revtex4-1}

\usepackage{graphicx}
\usepackage[utf8]{inputenc}
\usepackage{bm}
\usepackage{epsf}
\usepackage{rotating}
\usepackage{booktabs}
\usepackage{epsfig,graphics,rotate,color}
\usepackage{wrapfig}
\usepackage{amssymb}
\usepackage{amsmath}
\usepackage{amsfonts}
\usepackage{array,hhline,dcolumn}%
\usepackage{gensymb}
\usepackage{placeins}
\usepackage{tabularx}
\usepackage{nameref}
\usepackage{orcidlink}
\usepackage{xcolor}

\usepackage[colorlinks=true,citecolor=blue,linkcolor=blue]{hyperref}
\usepackage[capitalise]{cleveref}
\usepackage{cancel}
\usepackage{textcomp}
\usepackage{cellspace}
\usepackage{multirow, makecell}
\usepackage[T5,T1]{fontenc}

\definecolor{newred}{RGB}{2.329,53,43}
\definecolor{newblue}{RGB}{74,125,185}

\usepackage{hyperref}

\hypersetup
{
    colorlinks=true,
    linkcolor={newred},
    urlcolor={newblue},
    filecolor={newred},
    citecolor={newblue},
}

\newcommand\tywin{AT2019fdr}
\newcommand\bran{AT2019dsg}

\begin{document}

\author{Simeon Reusch \orcidlink{0000-0002-7788-628X}}
\affiliation{Deutsches Elektronen Synchrotron DESY, Platanenallee 6, D-15738 Zeuthen, Germany}
\affiliation{Institut f\"ur Physik, Humboldt-Universit\"at zu Berlin, D-12489 Berlin, Germany}

\author{Robert Stein \orcidlink{0000-0003-2434-0387}}
\affiliation{Deutsches Elektronen Synchrotron DESY, Platanenallee 6, D-15738 Zeuthen, Germany}
\affiliation{Institut f\"ur Physik, Humboldt-Universit\"at zu Berlin, D-12489 Berlin, Germany}
\affiliation{Division of Physics, Mathematics, and Astronomy, California Institute of Technology, 1200 East California Boulevard, MC 249-17, Pasadena, CA 91125, USA}

\author{Marek Kowalski \orcidlink{0000-0001-8594-8666}}
\affiliation{Deutsches Elektronen Synchrotron DESY, Platanenallee 6, D-15738 Zeuthen, Germany}
\affiliation{Institut f\"ur Physik, Humboldt-Universit\"at zu Berlin, D-12489 Berlin, Germany}

\author{Sjoert van Velzen \orcidlink{0000-0002-3859-8074}}
\affiliation{Leiden Observatory, Leiden University, Postbus 9513, 2300 RA, Leiden, Netherlands}

\author{Anna Franckowiak \orcidlink{0000-0002-5605-2219}}
\affiliation{Deutsches Elektronen Synchrotron DESY, Platanenallee 6, D-15738 Zeuthen, Germany}
\affiliation{Fakult\"at f\"ur Physik \& Astronomie, Ruhr-Universit\"at Bochum, D-44780 Bochum, Germany}

\author{Cecilia Lunardini \orcidlink{0000-0002-9253-1663}}
\affiliation{Department of Physics, Arizona State University, Tempe, AZ 85287-1504, USA}

\author{Kohta Murase \orcidlink{0000-0002-5358-5642}}
\affiliation{Department of Physics; Department of Astronomy and Astrophysics; Center for Multimessenger Astrophysics, Institute for Gravitation and the Cosmos, The Pennsylvania State University, University Park, Pennsylvania 16802, USA}
\affiliation{Center for Gravitational Physics, Yukawa Institute for Theoretical Physics, Kyoto University, Kyoto, Kyoto 606-8502, Japan}

\author{Walter Winter\orcidlink{0000-0001-7062-0289}}
\affiliation{Deutsches Elektronen Synchrotron DESY, Platanenallee 6, D-15738 Zeuthen, Germany}

\author{James C. A. Miller-Jones \orcidlink{0000-0003-3124-2814}}
\affiliation{International Centre for Radio Astronomy Research -- Curtin University, GPO Box U1987, Perth, WA 6845, Australia}

\author{Mansi M. Kasliwal \orcidlink{0000-0002-5619-4938}}
\affiliation{Division of Physics, Mathematics, and Astronomy, California Institute of Technology, 1200 East California Boulevard, MC 249-17, Pasadena, CA 91125, USA}

\author{Marat Gilfanov \orcidlink{0000-0003-4029-6769}}
\affiliation{Space Research Institute (IKI), Russian Academy of Sciences, Profsoyuznaya ul. 84/32, Moscow, 117997 Russia}
\affiliation{Max-Planck-Institut f\"ur Astrophysik, Karl-Schwarzschild-Straße 1, D-85741 Garching, Germany}

\author{Simone Garrappa \orcidlink{0000-0003-2403-4582}}
\affiliation{Deutsches Elektronen Synchrotron DESY, Platanenallee 6, D-15738 Zeuthen, Germany}
\affiliation{Institut f\"ur Physik, Humboldt-Universit\"at zu Berlin, D-12489 Berlin, Germany}

\author{Vaidehi S. Paliya \orcidlink{0000-0001-7774-5308}}
\affiliation{Aryabhatta Research Institute of Observational Sciences (ARIES), Manora Peak, Nainital-263001, Uttarakhand, India}

\author{Tom\'as Ahumada \orcidlink{0000-0002-2184-6430}}
\affiliation{Department of Astronomy, University of Maryland, College Park, MD 20742, USA}

\author{Shreya Anand \orcidlink{0000-0003-3768-7515}}
\affiliation{Division of Physics, Mathematics, and Astronomy, California Institute of Technology, 1200 East California Boulevard, MC 249-17, Pasadena, CA 91125, USA}

\author{Cristina Barbarino}
\affiliation{Department of Astronomy, The Oskar Klein Centre, Stockholm University: SE-106 91 Stockholm, Sweden}

\author{Eric C. Bellm \orcidlink{0000-0001-8018-5348}}
\affiliation{DIRAC Institute, Department of Astronomy, University of Washington, 3910 15th Avenue NE, Seattle, WA 98195, USA}

\author{Valéry Brinnel}
\affiliation{Institut f\"ur Physik, Humboldt-Universit\"at zu Berlin, D-12489 Berlin, Germany}

\author{Sara Buson \orcidlink{0000-0002-3308-324X}}
\affiliation{Institut f\"ur Theoretische Physik and Astrophysik, Universit\"at W\"urzburg, D-97074 W\"urzburg, Germany}

\author{S. Bradley Cenko \orcidlink{0000-0003-1673-970X}}
\affiliation{Astrophysics Science Division, NASA Goddard Space Flight Center, 8800 Greenbelt Road, Greenbelt, MD 20771, USA}
\affiliation{Joint Space-Science Institute, University of Maryland, College Park, MD 20742, USA}

\author{Michael W. Coughlin \orcidlink{0000-0002-8262-2924}}
\affiliation{School of Physics and Astronomy, University of Minnesota, Minneapolis, MN 55455, USA}

\author{Kishalay De \orcidlink{0000-0002-8989-0542}}
\affiliation{Division of Physics, Mathematics, and Astronomy, California Institute of Technology, 1200 East California Boulevard, MC 249-17, Pasadena, CA 91125, USA}

\author{Richard Dekany \orcidlink{0000-0002-5884-7867}}
\affiliation{Caltech Optical Observatories, California Institute of Technology, 1200 East California Boulevard, MC 249-17, Pasadena, CA 91125, USA}

\author{Sara Frederick \orcidlink{0000-0001-9676-730X}}
\affiliation{Department of Astronomy, University of Maryland, College Park, MD 20742, USA}

\author{Avishay Gal-Yam \orcidlink{0000-0002-3653-5598}}
\affiliation{Department of Particle Physics and Astrophysics, Weizmann Institute of Science, 234 Herzl Street 76100 Rehovot, Israel}

\author{Suvi Gezari \orcidlink{0000-0003-3703-5154}}
\affiliation{Space Telescope Science Institute, 3700 San Martin Dr., Baltimore, MD 21218, USA}
\affiliation{Department of Astronomy, University of Maryland, College Park, MD 20742, USA}

\author{Marcello Giroletti \orcidlink{0000-0002-8657-8852}}
\affiliation{INAF - Osservatorio di Astrofisica e Scienza dello Spazio di Bologna, via Gobetti 93/3, I-40129 Bologna, Italy}

\author{Matthew J. Graham \orcidlink{0000-0002-3168-0139}}
\affiliation{Division of Physics, Mathematics, and Astronomy, California Institute of Technology, 1200 East California Boulevard, MC 249-17, Pasadena, CA 91125, USA}

\author{Viraj Karambelkar \orcidlink{0000-0003-2758-159X}}
\affiliation{Division of Physics, Mathematics, and Astronomy, California Institute of Technology, 1200 East California Boulevard, MC 249-17, Pasadena, CA 91125, USA}

\author{Shigeo S. Kimura \orcidlink{0000-0003-2579-7266}}
\affiliation{Frontier Research Institute for Interdisciplinary Sciences; Astronomical Institute, Tohoku University, Sendai 980-8574 Japan}

\author{Albert K. H. Kong \orcidlink{0000-0002-5105-344X}}
\affiliation{Institute of Astronomy, National Tsing Hua University, No. 101 Section 2 Kuang-Fu Road, Hsinchu 30013, Taiwan}

\author{Erik C. Kool \orcidlink{0000-0002-7252-3877}}
\affiliation{Department of Astronomy, The Oskar Klein Centre, Stockholm University: SE-106 91 Stockholm, Sweden}

\author{Russ R. Laher \orcidlink{0000-0003-2451-5482}}
\affiliation{IPAC, California Institute of Technology, 1200 East California Boulevard, MC 249-17, Pasadena, CA 91125, USA}

\author{Pavel Medvedev \orcidlink{0000-0002-9380-8708}}
\affiliation{Space Research Institute (IKI), Russian Academy of Sciences, Profsoyuznaya ul. 84/32, Moscow, 117997 Russia}

\author{Jannis Necker \orcidlink{0000-0003-0280-7484}}
\affiliation{Deutsches Elektronen Synchrotron DESY, Platanenallee 6, D-15738 Zeuthen, Germany}
\affiliation{Institut f\"ur Physik, Humboldt-Universit\"at zu Berlin, D-12489 Berlin, Germany}

\author{Jakob Nordin \orcidlink{0000-0001-8342-6274}}
\affiliation{Institut f\"ur Physik, Humboldt-Universit\"at zu Berlin, D-12489 Berlin, Germany}

\author{Daniel A. Perley \orcidlink{0000-0001-8472-1996}}
\affiliation{Astrophysics Research Institute, Liverpool John Moores University, 146 Brownlow Hill, Liverpool L3 5RF, UK}

\author{Mickael Rigault \orcidlink{0000-0002-8121-2560}}
\affiliation{Univ Lyon, Univ Claude Bernard Lyon 1, CNRS/IN2P3, UMR 5822, F-69622, Villeurbanne, France}

\author{Ben Rusholme \orcidlink{0000-0001-7648-4142}}
\affiliation{IPAC, California Institute of Technology, 1200 East California Boulevard, MC 249-17, Pasadena, CA 91125, USA}

\author{Steve Schulze \orcidlink{0000-0001-6797-1889}}
\affiliation{The Oskar Klein Centre, Physics Department of Physics, Stockholm University, Albanova University Center, SE-106 91 Stockholm, Sweden}

\author{Tassilo Schweyer \orcidlink{0000-0001-8948-3456}}
\affiliation{Department of Astronomy, The Oskar Klein Centre, Stockholm University: SE-106 91 Stockholm, Sweden}

\author{Leo P. Singer \orcidlink{0000-0001-9898-5597}}
\affiliation{NASA Goddard Space Flight Center, University of Maryland, Baltimore County, Greenbelt, MD 20771, USA}

\author{Jesper Sollerman \orcidlink{0000-0003-1546-6615}}
\affiliation{Department of Astronomy, The Oskar Klein Centre, Stockholm University: SE-106 91 Stockholm, Sweden}

\author{Nora Linn Strotjohann \orcidlink{0000-0002-4667-6730}}
\affiliation{Department of Particle Physics and Astrophysics, Weizmann Institute of Science, 234 Herzl Street 76100 Rehovot, Israel}

\author{Rashid Sunyaev \orcidlink{0000-0002-2764-7192}}
\affiliation{Space Research Institute (IKI), Russian Academy of Sciences, Profsoyuznaya ul. 84/32, Moscow, 117997 Russia}
\affiliation{Max-Planck-Institut f\"ur Astrophysik, Karl-Schwarzschild-Straße 1, D-85741 Garching, Germany}

\author{Jakob van Santen \orcidlink{0000-0002-2412-9728}}
\affiliation{Deutsches Elektronen Synchrotron DESY, Platanenallee 6, D-15738 Zeuthen, Germany}

\author{Richard Walters}
\affiliation{Caltech Optical Observatories, California Institute of Technology, 1200 East California Boulevard, MC 249-17, Pasadena, CA 91125, USA}

\author{B. Theodore Zhang \orcidlink{0000-0003-2478-333X}}
\affiliation{Center for Gravitational Physics, Yukawa Institute for Theoretical Physics, Kyoto University, Kyoto, Kyoto 606-8502, Japan}

\author{Erez Zimmerman \orcidlink{0000-0001-8985-2493}}
\affiliation{Department of Particle Physics and Astrophysics, Weizmann Institute of Science, 234 Herzl Street 76100 Rehovot, Israel}

\title{Candidate Tidal Disruption Event \tywin{} Coincident with a High-Energy Neutrino}

\begin{abstract}
\newpage
The origins of the high-energy cosmic neutrino flux remain largely unknown. Recently, one high-energy neutrino was associated with a tidal disruption event (TDE). Here we present \tywin{}, an exceptionally luminous TDE candidate, coincident with another high-energy neutrino. Our observations, including a bright dust echo and soft late-time X-ray emission, further support a TDE origin of this flare. The probability of finding two such bright events by chance is just 0.034\%. We evaluate several models for neutrino production and show that \tywin{} is capable of producing the observed high-energy neutrino, reinforcing the case for TDEs as neutrino sources.
\end{abstract}

\maketitle
Neutrino astronomy is at a crossroads: While a flux of high-energy cosmic neutrinos has been firmly established through observations with the IceCube Neutrino Observatory \cite{icecube_detector, Aartsen:2015rwa, 2014PhRvL.113j1101A, high_energy_flux}, identifying their sources has been a challenge. The emission of cosmic neutrinos is a smoking-gun signature for hadronic acceleration (see \cite{ahlers_18} for a recent review), and discovering their sources will allow us to resolve long-standing questions about the production sites of high-energy cosmic rays.

Three sources have thus far been associated with neutrinos at post-trial significance of $\approx 3\sigma$, which can be considered evidence for a true association \cite{3sigma}. In 2017, the flaring blazar TXS 0506+056 was identified as the likely source of neutrino alert IC170922A \cite{IceCube:2018dnn}. This same source was also associated with a neutrino flare in 2014-15 \cite{ic_txs_ps_18}, occurring during a period without significant electromagnetic flaring activity \cite{garrappa_19}. 
In 2019, the Tidal Disruption Event (TDE) \bran{} was identified as the likely source of IC191001A \cite{Bran}. More recently, the IceCube collaboration reported a clustering of neutrinos from the direction of the Active Galactic Nucleus (AGN) in the nearby galaxy NGC 1068 \cite{Aartsen:2019fau}. AGN are galaxies with high levels of supermassive black hole (SMBH) accretion, and have been long proposed as high-energy neutrino sources \cite{Ber77,Eichler:1979yy,stecker,1993A&A...269...67M,SZABO1994375,Murase_2017}. These associations and other conceptual arguments suggest that the neutrino flux may arise from a mixture of different astrophysical populations \cite{Murase:2015xka,Palladino:2018evm,pie-chart}, although AGN or another source class can still be dominant \cite{murase_seyfert}.

TDEs are rare transients that occur when stars pass close enough to SMBHs and get destroyed by tidal forces. The result of this destruction is a luminous electromagnetic flare with a timescale of $\sim$ months. Theoretical studies have suggested that TDEs might be sources of high-energy neutrinos and ultrahigh-energy cosmic rays \cite{2009ApJ...693..329F,Murase:2008zzc,Wang:2011ip,farrar2014tidal,2016PhRvD..93h3005W,2017MNRAS.469.1354D,2016PhRvD..93h3005W,senno_TDE,Guepin:2017abw,lunardini_TDE,dai_TDE,Zhang:2017hom,Biehl:2017hnb,Guepin:2017abw,2019ApJ...886..114H, Winter:2020ptf,murase_bran, 2020PhRvD.102h3028L,fang_20, Winter:2021lyo}. Some models consider emission from a relativistic jet, while others propose additional neutrino production scenarios e.g., in a disk, disk corona, or wind (see \citet{Hayasaki:2021jem} for a review). 
In the case of \bran{}, radio observations confirmed long-lived non-thermal emission from the source \citep{Bran, Cannizzaro:2020xzc, Cendes:2021bvp, mohan_21, matsumoto_21}, but generally disfavor those models relying on the presence of an on-axis relativistic jet \citep{Winter:2020ptf} in the standard leptonic radio emission scenario.

TDEs and AGN flares are ultimately both modes of SMBH accretion. Some models highlight this potential similarity, and have developed common frameworks for neutrino emission from both cases (see e.g. \cite{murase_bran}). However, AGN flares are vastly more numerous than TDEs, injecting significantly more energy into the universe. If TDEs nonetheless contribute significantly to the neutrino flux, they must be very efficient neutrino emitters. Whether there are particular characteristics of TDEs that enable efficient neutrino production, and whether these conditions are also present in particular classes of AGN accretion flares, remain open questions for neutrino astronomy.

Bridging these two astrophysical populations, we here report new observations of \tywin{}, a candidate TDE in a Narrow-Line Seyfert 1 (NLSy1) active galaxy \citep{sara}. Similar to \bran{}, \tywin{} was identified as a likely neutrino source by the neutrino follow-up program of the Zwicky Transient Facility (ZTF) \citep{ztf_overview, ztf_science_objectives, ztf_instrument}. \tywin{} lies within the reported 90\% localization region of the IceCube high-energy neutrino IC200530A \cite{IC200530A}. The observations were processed by \texttt{nuztf}, our multi-messenger analysis pipeline \citep{nuztf, ampel}, which searches for extragalactic transients in spatial and temporal coincidence with high-energy neutrinos \cite{2007APh....27..533K, Bran}, and \tywin{} was reported as a candidate \cite{ztf_IC200530A}.

\begin{figure*}
\centering
\includegraphics[width=1.0\textwidth]{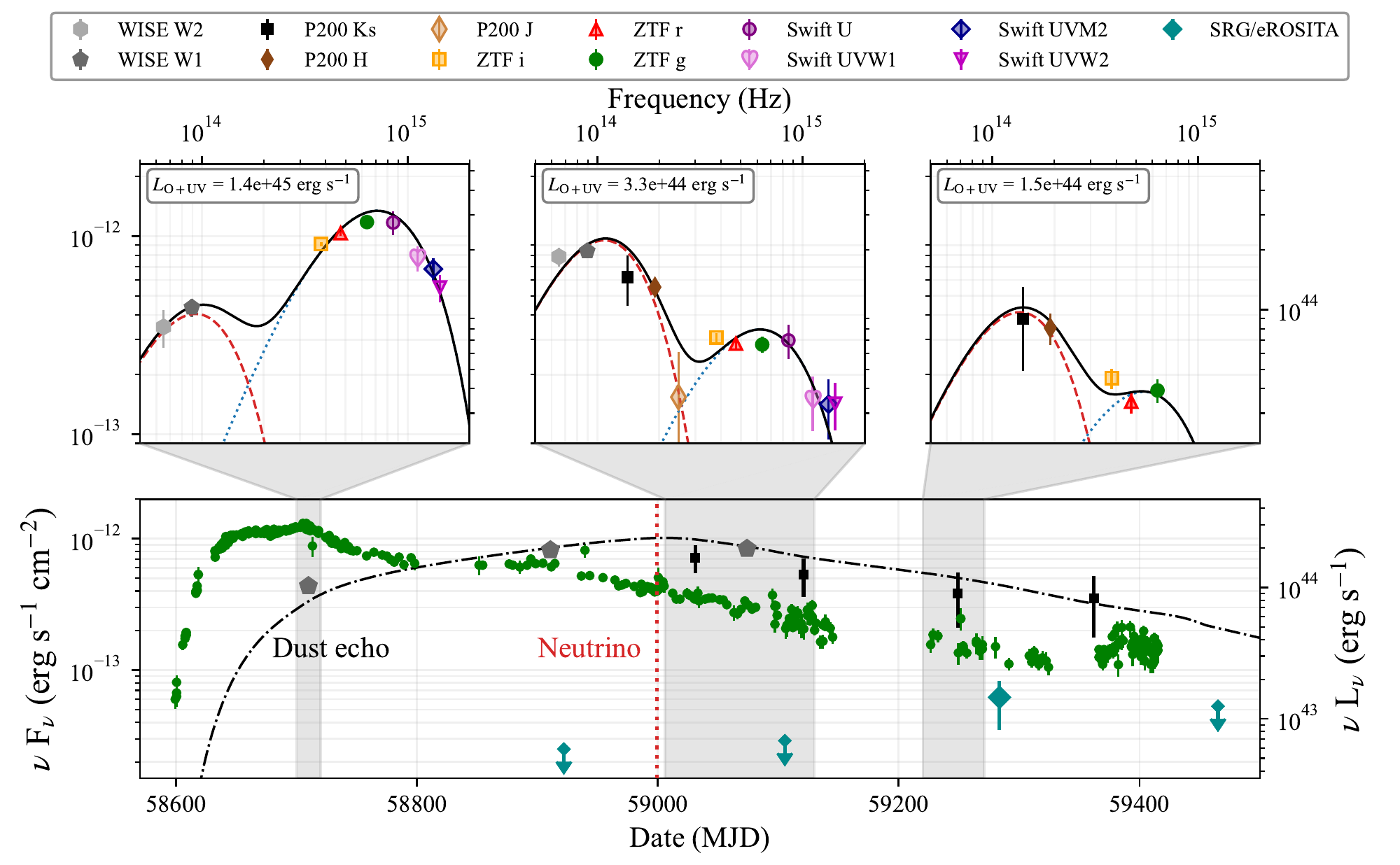}
\caption{The bottom plot shows the lightcurve in the optical ZTF g-band, the infrared P200 Ks- and  \textit{WISE} W1-band as well as the modeled dust echo (black line, dashdot), with the neutrino arrival time marked with a red dotted vertical line. The \textit{SRG}/eROSITA X-ray measurements are also included. The shaded gray areas are averaged and their respective SEDs are shown in the top panels, including a fitted blue and a red blackbody (blue dashed and red dotted curve; lab frame), as well as the combined spectrum (black solid curve). The left axes all show $\nu F_\nu$, where $F_\nu$ is the spectral flux density at frequency $\nu$, while the right axes show $\nu L_\nu$, where $L_\nu$ is the luminosity at frequency $\nu$. The second epoch (middle plot on top) encompasses several months to include both \textit{WISE} and P200 infrared data points. The global values for line-of-sight dust extinction are $A_V=0.45^{+0.14}_{-0.14}$ mag, assuming  $R_V=3.1$ and the Calzetti attenuation law \citep{calzetti}. Note that the X-ray measurements were not included in the blackbody fits. The luminosities are given in the source rest frame.}
\label{fig:lightcurve_and_sed}
\end{figure*}

\tywin{}, a long-duration flare (see Fig. \ref{fig:lightcurve_and_sed}) of apparent nuclear origin, was first discovered by ZTF one year prior to the neutrino detection \cite{sara, tywin_discovery}. \tywin{} reached a peak flux of $1.3 \times 10^{-12}~{\rm erg}~{\rm s}^{-1}~{\rm cm}^{-2}$ in the optical ZTF g-band on 2019, August 10, before slowly fading. With a peak g-band luminosity of $L_{\rm peak} = 2.9 \times 10^{44}$ erg s$^{-1}$, \tywin{} was an extraordinarily luminous event. At the time of neutrino detection, it had decayed to $\sim 30$\% of its peak flux, and was still detected by ZTF as of August 2021. Forced photometry using data from ZTF (up to 400 days prior to the flare) as well as from the Palomar Transient Facility (2010--2016) \cite{ptf} shows no historical variability.

\tywin{} was classified as a probable TDE, though an extreme AGN flare origin could not be ruled out \citep{sara}. High-resolution spectra yielded a redshift of $z=0.267$. Using a spectrum from the Alhambra Faint Object Spectrograph and Camera (ALFOSC), on the Nordic Optical Telescope (NOT; PI: Sollerman), a virial black hole mass estimate of $M_{\rm BH} = 10^{7.55 \pm 0.13} M_\odot$ was inferred; for further details refer to the Supplemental Material (SM).

Though the classification of \tywin{} based on early observations included the possibility of it being a Type II superluminous supernova (SLSN-II, see \cite{avishay_slsn}) \cite{2019TNSCR1016....1C}, leading to further studies \cite{tywin_slsn}, its subsequent spectroscopic and photometric evolution was not consistent with expectations for SLSNe.
\citet{sara} already disfavored the SLSN hypothesis based on the long-lived U-band and the UV emission, the flare's longevity, emission at the blue end of the Balmer line profiles as well as its proximity to the nucleus of the galaxy. Here we add a late-time X-ray detection and the detection of a strong infrared echo, rendering a SLSN interpretation less likely (see below).

After discovery, \tywin{} was also observed by the Ultraviolet and Optical Telescope (UVOT) \citep{swift_uvot} aboard the \textit{Neil Gehrels Swift Observatory} (\textit{Swift}) \citep{swift, sara}. Additional observations continued up to 2020, June 7, including one epoch shortly after the neutrino detection. By that point, the transient had faded by 84\% in the UVW1-band from its peak luminosity of $2.1 \times 10^{44}$ erg s$^{-1}$. \tywin{} was not detected in any of the simultaneous X-ray observations by the \textit{Swift} X-ray Telescope (XRT) \citep{swift_xrt}, yielding a combined $3\sigma$ flux upper limit of $1.4\times 10^{-13}$ erg s$^{-1}$ cm$^{-2}$ for all observations before neutrino arrival (corrected for absorption).

The position of \tywin{} was also visited by the eROSITA telescope \cite{erosita} aboard the \textit{Spectrum-Roentgen-Gamma} (\textit{SRG}) mission \cite{erosita_xrg} four times. The first two visits did not detect an excess, with a mean flux upper limit of $2.7\times 10^{-14}$ erg s$^{-1}$ cm$^{-2}$ at the 95\% confidence level. However, at the third visit on 2021, March 10--11, it detected late time X-ray emission from the transient with an energy flux of $6.2_{-2.1}^{+2.7}\times 10^{-14}$~erg s$^{-1}$ cm$^{-2}$ in the 0.3--2.0\,keV band, thus showing temporal evolution in the X-ray flux (see Fig. \ref{fig:lightcurve_and_sed}). The detection displayed a very soft thermal spectrum with a best fit blackbody temperature of $56_{-26}^{+32}$ eV. 

The softness of the spectrum provides further evidence for \tywin{} being a TDE rather than regular AGN variability, where soft spectra are rare \cite{tde_xray}. Though NLSy1 galaxies generally exhibit softer X-ray spectra, the temperature of \tywin{} is atypically low even in this context (lower than all NLSy1s in \cite{1999ApJS..125..317L} and \cite{10.1046/j.1365-8711.1999.02811.x}). Furthermore, X-ray emission is rarely seen for SLSNe \cite{2018ApJ...864...45M}, with only the first SLSN ever observed, SCP 06F6 \cite{Barbary:2008ge}, possibly showing an X-ray flux exceeding the luminosity of \tywin{} \cite{2009ApJ...697L.129G}. This provides more evidence against the SLSN classification.

\tywin{} was further detected at mid-infrared (MIR) wavelengths as part of routine NEOWISE survey observations \cite{NEOWISE} by the Wide-field Infrared Survey Explorer (\textit{WISE}) \cite{WISE}. Using pre-flare archival NEOWISE data as baseline, a substantial flux increase was detected in both W1- and W2-band. MIR emission reached a peak luminosity of $1.9\times 10^{44} \rm\,erg\,s^{-1}$ on 2020, August 13, over one year after the optical/UV peak. Complementary near-infrared (NIR) measurements were taken with the P200 Wide Field Infrared Camera (WIRC, \citep{wirc}) in the J-, H- and Ks band. After subtracting a synthetic host model (see SM), a fading transient infrared signal was detected in all three bands; see Fig. \ref{fig:lightcurve_and_sed}.

We modeled this lightcurve as a composite of two unmodified blackbodies (a `blue' and a `red' blackbody). We interpret the time-delayed infrared emission as a dust echo: The blue blackbody heats surrounding dust, which then starts to glow. The lightcurve of this dust echo was inferred using the method described in \cite{dustecho} and the corresponding fit is shown in Fig. \ref{fig:lightcurve_and_sed}. An optical/UV bolometric luminosity of $L=1.4^{+0.1}_{-0.1} \times 10^{45}$ erg s$^{-1}$ at peak was derived. By integrating this component over time, we derived a total bolometric energy of $E_{\rm bol} = 3.4 \times 10^{52}\rm\,erg$ (the red blackbody was not added, as dust absorption is already accounted for through the extinction correction). This is almost twice the inferred bolometric energy of ASASSN-15lh, which was one of the brightest transients ever reported \cite{asassn15lh} and was suggested to be a TDE \cite{asassn_15lh_tde}. Furthermore, the energy budget, bolometric evolution and luminous dust echo suggest that \tywin{} belongs to a class of TDE candidates observed in AGN (similar to PS1-10adi \cite{ps1-adi}, AT2017gbl \cite{at2017gbl} or Arp 299-B AT1 \cite{Arp299-BAT1}). For details on the modeling methods, see SM.

Following the neutrino detection, we performed radio observations of \tywin{} with a dedicated Very Large Array (VLA) \citep{vla_11} Director’s Discretionary Time (DDT) program (PI: Stein) three times over a period of four months, and obtained multi-frequency detections. \tywin{} shows a featureless power law spectrum consistent with optically thin synchrotron emission above $\sim$ 1 GHz with no significant intrinsic evolution between the epochs (see SM). The peak flux density was $0.39 \pm 0.03$ mJy in the 1--2\,GHz band. The lack of apparent evolution suggests that the radio emission is not related to the transient, but rather originated from the AGN host. An additional sub-dominant transient component could be present.

No gamma rays were detected by the \textit{Fermi} Large Area Telescope (\textit{Fermi}-LAT) \citep{fermi} between the first detection of \tywin{} and one year after neutrino detection, yielding an upper limit of $1.3 \times 10^{-12} \, \rm erg\, s^{-1}\,cm^{-2}$ (see \cite{lancel}).

\tywin{} is the second probable neutrino--TDE association found by ZTF. To calculate the probability of finding two such coincident events by chance, while accounting for the fact that some TDEs will not be spectroscopically classified, we developed a broader sample of photometrically-selected `candidate TDEs'. We selected `nuclear' transients that are at least as bright as \tywin{} from the sample of ZTF transients, and applied cuts to identify TDE-like rise- and decay-times (see SM and \cite{lancel} for details). Our sample begins in 2018 (the ZTF survey start), and we further required a flare peak date before July 2020. We excluded only transients for which a TDE origin was ruled out through spectroscopic classification (i.e. our sample contains all unclassified candidates and all classified TDEs). To compute the sky source density at any given time, we conservatively estimated their average lifetime at 1 year after discovery, yielding an effective source density of $1.7 \times 10^{-4}$ per deg$^{2}$ of sky in the ZTF footprint (most TDEs evolve on shorter timescales, which -- if accounted for -- would reduce the effective source density). When including all 24 neutrinos followed up by our program by September 2021 (covering a combined area of 154.33 deg$^{2}$, see SM), the probability of finding any photometrically-selected TDE candidate by chance is $2.6$ $\times 10^{-2}$, while the probability of finding two by chance is $3.4 \times 10^{-4}$ (3.4$\sigma$). We emphasize that these estimates rely solely on the optical flux and a nuclear location in the host galaxy, and thus do not account for the additional luminous dust echoes or post-flare X-ray detections observed for \bran{} and \tywin{}.

\textbf{\textit{Neutrino emission from AT2019fdr}}: With a single neutrino observed in association with \tywin{}, the inference of the neutrino flux will be subject to a large Eddington bias~\cite{nora_eddington} and hence very uncertain. However, we can make a more robust statement on the neutrino flux by considering the underlying population (see e.g.\ \cite{Bran}). The detection of two high-energy neutrinos implies a mean expectation for the full TDE catalog in the range $0.36 < N_{\nu, \textup{tot}} < 6.30$ at 90\% confidence, where $N_{\nu, \textup{tot}}$ is the cumulative neutrino expectation for the nuclear transients that ZTF has observed.
\tywin{} emits $\sim 2\%$ of the g-band peak energy flux for the population of nuclear transients, consisting of the 17 published ZTF TDEs (see \cite{winterishere}) and all TDE candidates as bright as \tywin{} (see SM for the latter). If we take this as a proxy for \tywin{}'s contribution to the neutrino emission, we would expect a total number of neutrinos $0.007 \lesssim N_{\nu} \lesssim 0.13$ for this source.

\begin{figure}
    \centering
    \includegraphics[width=1\columnwidth]{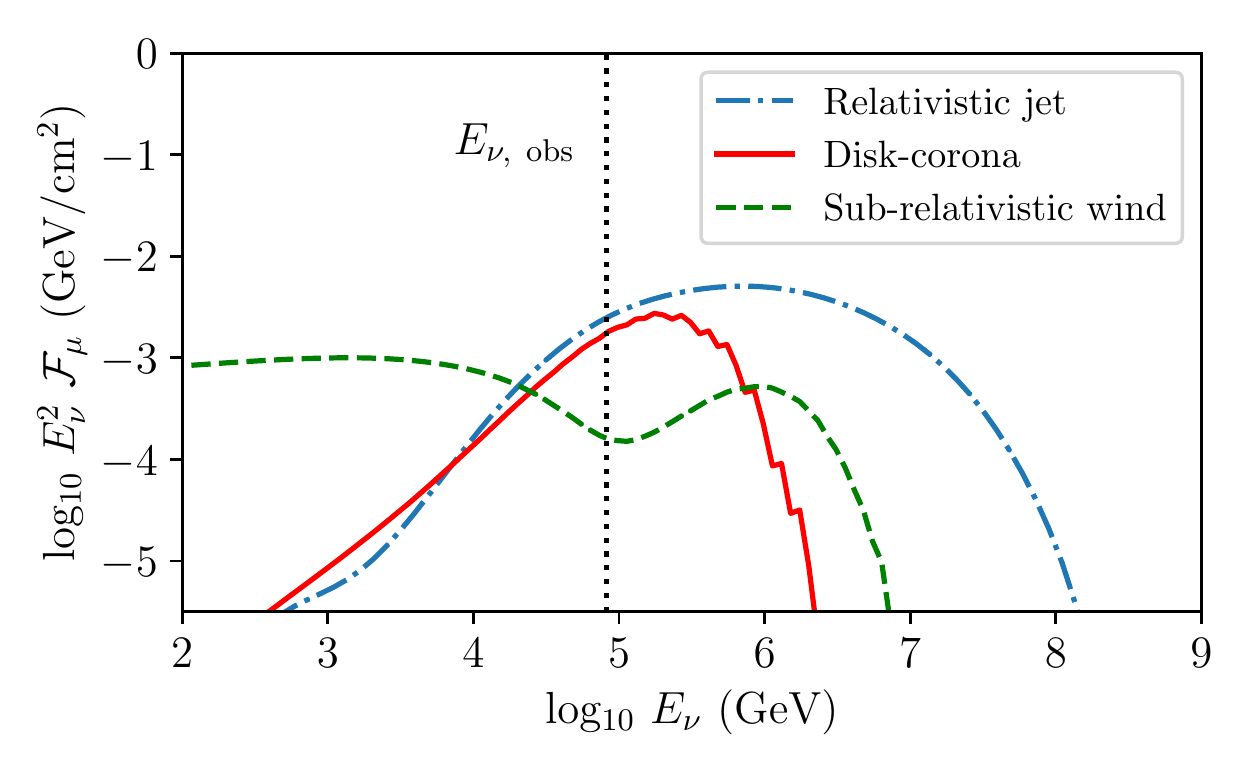}
    \caption{Neutrino fluence for the three models described here. The reported energy of the neutrino event \cite{IC200530A}, represented by the dotted vertical line, should be viewed as a lower limit to the neutrino energy.}
    \label{fig:fluence}
\end{figure}

This estimate can be compared to model expectations. We present three different models invoking $p\gamma$ and/or $pp$ interactions, where protons are efficiently accelerated in a disk corona, a sub-relativistic wind or a relativistic jet (see SM). The resulting spectra are shown in Fig.\ \ref{fig:fluence}. All models can explain the observed energy of the IC200530A neutrino event; they also make predictions for the underlying `neutrino lightcurve', though this can only be resolved once many neutrinos from TDEs have been detected. The obtained neutrino luminosities $L_\nu \lesssim 0.1 \, L_{\mathrm{Edd}} \simeq 5 \times 10^{44}\rm\,erg\,s^{-1}$ are consistent with theoretical expectations for most models~\cite{Winter:2021lyo}.

In accretion flow models \cite{murase_bran,2019ApJ...886..114H}, the virial theorem implies a cosmic ray acceleration efficiency $\eta_{\rm CR}< (1/40){(R/10\,R_{\rm S})}^{-1}$~\cite{murase_bran} for a cosmic-ray luminosity $L_{\rm CR}=\eta_{\rm CR}\dot{M} c^2$, where $R$ is the emission radius and $R_{\rm S}=2\,GM/c^2$ is the Schwarzschild radius. Even for a mass accretion rate of $\dot{M}\sim 10\,L_{\rm Edd}/c^2$, the neutrino luminosity would not exceed $\sim10^{44}\rm\,erg\,s^{-1}$. 
In the case of \tywin{}, the Eddington ratio $\lambda_{\rm Edd}\equiv L/L_{\rm Edd}\lesssim0.07-0.3$ in the first 2 epochs, implying accretion near the Eddington limit around the peak and sub-Eddington accretion around the time of the neutrino detection. For such high accretion the disk plasma is collisional, while the coronal region may allow particle acceleration and non-thermal neutrino production~\cite{murase_bran}. This model yields $N_{\nu}\sim0.007$ when evaluating its spectrum under the effective area of the neutrino alert channel \cite{GFU_EHE}. This is within the expected range, albeit at its lower end. The time delay is consistent with quasi-steady coronal emission. Alternatively, because the accretion rate gradually decreases, the neutrino time delay can be attributed to the formation of a collisionless corona that allows ion acceleration~\cite{murase_bran}.

We also considered a sub-relativistic wind with a velocity of $\sim0.1c$, consistent with what was observed for \bran{}. Such a wind is naturally launched from the TDE disk (e.g.,~\cite{Jiang+19}), and may interact with tidal disruption debris. A strong shock is also expected from interactions between tidal streams. Ions can be accelerated at the shock via diffusive shock acceleration and produce neutrinos through inelastic $pp$ and $p\gamma$ collisions~\cite{murase_bran}. In this sub-relativistic wind model, the maximum proton energy can be as high as $\sim10-100$~PeV. If the cosmic ray luminosity is three times the optical luminosity, the expected number of muon neutrinos is $N_{\nu}\sim0.002$, which falls outside the empirical range for this baryon loading factor. The neutrino light curve would trace the wind luminosity in the calorimetric limit, and the time delay is consistent with quasi-steady radio emission.

In the relativistic jet model, external target photons from the disk are back-scattered into the jet frame. Here we followed \cite{Winter:2020ptf} for \bran{}, but adopted a unified model~\cite{Mummery:2021nqy} to extrapolate to higher SMBH masses as given for \tywin{}. We estimated a thermal far UV to X-ray spectrum with $T \simeq 34 \, \mathrm{eV}$. This turned out to be consistent with the late-time X-ray detection within the uncertainties. The isotropization timescale of the photons is expected to be given by the system size, suggesting a possible correlation with the dust echo; as a consequence the isotropized X-ray and dust echo lightcurves look very similar. The jet model allows for efficient particle acceleration and results in a relatively large number of 0.027 neutrino events with a maximum $L_\nu \simeq 0.05 \, L_{\mathrm{Edd}}$ thanks to the beaming effect; however, direct signatures of the jet have not been observed.

\textbf{\textit{Conclusions}}:
\tywin{} was an exceptionally bright nuclear transient that was already identified in the literature as a probable TDE in an active galaxy \cite{sara}. In this work, we have presented new observational data, including the identification of a strong dust echo and soft late-time X-ray emission, which further support a TDE origin for this flare.

\tywin{} was a very long-lived transient, one of the most luminous ever detected. For a TDE, the energy release would require a very massive star \cite{Ryu:2020gxf}. However, unlike for TDEs in quiescent galaxies, the AGN disk in \tywin{} might provide the system with additional energy \cite{2019ApJ...881..113C}. Furthermore, the post star-burst nature of the host increases the expected rate for TDEs \cite{arcavi, french, 2018MNRAS.480.5060S}.

\tywin{} was the second candidate neutrino-TDE identified by our ZTF follow-up program. While \tywin{} was far more luminous than \bran{}, the first TDE associated with a high-energy neutrino, it was also more distant. As a result, the two objects have comparable bolometric energy fluxes. The probability for finding two such bright neutrino-coincident TDEs by chance is just $3.4 \times 10^{-4}$, a sevenfold decrease relative to the previously-reported single association \cite{Bran}.  The gain due to the second association is somewhat offset by the larger neutrino sample and the more inclusive candidate TDE selection. Within the framework of this paper, the association of a second object results in a reduction of the chance probability by a factor of 75 versus a single association.

\tywin{} and \bran{} share other similarities beyond their potential association with a high-energy neutrino. Intriguingly, \bran{} also displayed an unusually strong dust echo signal \cite{lancel}, indicating that the presence of large amounts of matter and an associated high star formation rate in the environment could be a common signature for high-energy neutrino production in such systems. A dedicated search for further associations based on this signature is presented in \cite{lancel} and provides more supporting evidence for neutrino production in TDEs.

We studied neutrino emission from \tywin{} using models previously applied to explain the observations of \bran{}. Similar to \bran{}, various plausible cosmic ray acceleration sites have been identified, such as the corona, a sub-relativistic wind, or a relativistic jet. The number of expected muon neutrinos predicted by the corona and jet models is consistent with empirical constraints derived from the two TDE-neutrino associations. All models require efficient neutrino production at a neutrino luminosity comparable to a fraction of the Eddington luminosity. The neutrino delay may be related to the size of the newly formed system (jet model) or the formation of a collisionless corona (corona model).

With two objects being associated with IceCube neutrino alerts, out of a number of 11.5 expected astrophysical neutrinos (summed alert signalness, see SM), we obtain a fraction of $18^{+38}_{-15}\%$ (90\% confidence level) of astrophysical neutrinos that could be explained due to ZTF-detected TDE candidates. Accounting for the incompleteness of our sample with the procedure in \citep{Bran}, our results imply that at least 7.8\% of astrophysical neutrinos would come from the broader TDE population.

The search for neutrinos resulting in public alerts has a high energy threshold to reduce the background. Even when considering the full energy range of IceCube \cite{ic_ps}, the expected number of neutrino events from \tywin{} remains below one. Therefore, the detection of additional lower-energy neutrinos from \tywin{} is not expected (see also the search by the ANTARES neutrino observatory \cite{antares}).

Fully understanding the role of TDEs as particle accelerators will only be possible with comprehensive multi-wavelength and -messenger data. While the detailed production processes remain uncertain, the observations presented here provide further evidence that TDEs are highly efficient sources of high-energy neutrinos. 

\nocite{tywincode, ampel, ztfquery, IC200530A, antares, gaia_dr2, lancel, hawc, integral, erosita_xrg, erosita, erosita_tde, hi4pi, sara, swift_procedure, 2009MNRAS.397.1177E, heasarc_webpimms, ztffps, ztflc, Breeveld2011a, wirc, De2020, 2006AJ....131.1163S, galfit, photutils, York02, vanVelzen20, Martin05, Million16, Stoughton02, ukidss, Conroy10, Foreman-Mackey14, WISE, NEOWISE, Jones15, Masci13, Bran, vlass, 2013ApJ...772...78B, 2012A&A...538A..81M, 2018ApJ...854...86E, 2013MNRAS.436.1258H, Generozov17, lmfit, extinctionpackage, calzetti, dustecho, VP06, Guo18, F99, dust, BG92, Shen11, Guo19, 2019ApJS..241...34S, mclure, ztf_alertsystem, ztf_dataprocessing, winterishere, IC190503A1, IC190503A2, IC190619A1, IC190619A2, IC190730A1, IC190730A2, IC190922B1, IC190922B2, IC190922B3, IC191001A1, IC191001A2, IC191001A3, IC200107A1, IC200107A2, IC200109A1, IC200109A2, IC200117A1, IC200117A2, IC200117A3, IC200512A1, IC200512A2, IC200530A1, IC200530A2, IC200530A3, IC200530A4, IC200620A1, IC200620A2, IC200916A1, IC200916A2, IC200916A3, IC200926A1, IC200926A2, IC200929A1, IC200929A2, IC201007A1, IC201007A2, IC201021A1, IC201021A2, IC201130A1, IC201130A2, IC201209A1, IC201209A2, IC201222A1, IC201222A2, IC210210A1, IC210210A2, IC210510A1, IC210510A2, IC210629A1, IC210629A2, IC210811A1, IC210811A2, IC210922A1, IC210922A2, murase_bran, 2019ApJ...886..114H, murase_seyfert, 2015ApJ...806..159K, McKinney+15,Jiang+19,Takeo+19, 2000CoPhC.124..290M, 2014ApJ...780...46I, 2014ApJ...784..169J, 2016ApJ...826...23T, Strubbe:2009qs,Miller:2015jda,Metzger:2015sea, Jiang+19, Winter:2020ptf, Dai:2018jbr, Mummery:2021otz,Mummery:2021qrc,Mummery:2021nqy,Mummery:2021rgp, Kochanek:2016zzg, Dai:2018jbr, Rees:1988bf,2012ApJ...760..103D, Murase:2014foa, mohan_21,matsumoto_21}

\bibliographystyle{apsrev4-2}
\bibliography{Candidate_Tidal_Disruption_Event_AT2019fdr_Coincident_with_a_High-Energy_Neutrino}

\section{Correspondence}
Correspondence and requests for materials should be addressed to Marek Kowalski (\href{mailto:marek.kowalski@desy.de}{marek.kowalski@desy.de}) and Simeon Reusch (\href{mailto:simeon.reusch@desy.de}{simeon.reusch@desy.de}).

\section{Acknowledgments}
S.R. acknowledges support by the Helmholtz Weizmann Research School on Multimessenger Astronomy, funded through the Initiative and Networking Fund of the Helmholtz Association, DESY, the Weizmann Institute, the Humboldt University of Berlin, and the University of Potsdam. A.F. acknowledges support by the Initiative and Networking Fund of the Helmholtz Association through the Young Investigator Group program (A.F.). C.L. acknowledges support from the National Science Foundation (NSF) with grant number PHY-2012195. 
The work of K.M. is supported by the NSF Grant No.~AST-1908689, No.~AST-2108466 and No.~AST-2108467, and KAKENHI No.~20H01901 and No.~20H05852. 
M.C. acknowledges support from the National Science Foundation with grant numbers PHY-2010970 and OAC-2117997. S.S. acknowledges support from the G.R.E.A.T research environment, funded by {\em Vetenskapsr\aa det}, the Swedish Research Council, project number 2016-06012. M.G., P.M. and R.S. acknowledge the partial support of this research by grant 19-12-00369 from the Russian Science Foundation. S.B. acknowledges financial support by the European Research Council for the ERC Starting grant \emph{MessMapp}, under contract no. 949555. A.G.Y.'s research is supported by the EU via ERC grant No. 725161, the ISF GW excellence center, an IMOS space infrastructure grant and BSF/Transformative and GIF grants, as well as The Benoziyo Endowment Fund for the Advancement of Science, the Deloro Institute for Advanced Research in Space and Optics, The Veronika A. Rabl Physics Discretionary Fund, Minerva, Yeda-Sela and the Schwartz/Reisman Collaborative Science Program; A.G.Y. is the incumbent of the The Arlyn Imberman Professorial Chair. E.C.K. acknowledges support from the G.R.E.A.T research environment funded by {\em Vetenskapsr\aa det}, the Swedish Research Council, under project number 2016-06012, and support from The Wenner-Gren Foundations. M.R. has received funding from the European Research Council (ERC) under the European Union's Horizon 2020 Research and Innovation Program (Grant Agreement No. 759194 - USNAC). N.L.S. is funded by the Deutsche Forschungsgemeinschaft (DFG, German Research Foundation) via the Walter Benjamin program – 461903330.

Based on observations obtained with the Samuel Oschin Telescope 48-inch and the 60-inch Telescope at the Palomar Observatory as part of the Zwicky Transient Facility project. ZTF is supported by the National Science Foundation under Grant No. AST-1440341 (until 2020 December 1) and No. AST-2034437 and a collaboration including Caltech, IPAC, the Weizmann Institute for Science, the Oskar Klein Center at Stockholm University, the University of Maryland, Deutsches Elektronen-Synchrotron and Humboldt University, the TANGO Consortium of Taiwan, the University of Wisconsin at Milwaukee, Trinity College Dublin, Lawrence Livermore National Laboratories, IN2P3, University of Warwick, Ruhr University Bochum and Northwestern University. Operations are conducted by COO, IPAC, and UW.

This work was supported by the GROWTH (Global Relay of Observatories Watching Transients Happen) project funded by the National Science Foundation Partnership in International Research and Education program under Grant No 1545949. GROWTH is a collaborative project between California Institute of Technology (USA), Pomona College (USA), San Diego State University (USA), Los Alamos National Laboratory (USA), University of Maryland College Park (USA), University of Wisconsin Milwaukee (USA), Tokyo Institute of Technology (Japan), National Central University (Taiwan), Indian Institute of Astrophysics (India), Inter-University Center for Astronomy and Astrophysics (India), Weizmann Institute of Science (Israel), The Oskar Klein Centre at Stockholm University (Sweden), Humboldt University (Germany).

This work is based on observations with the eROSITA telescope on board the SRG observatory. The SRG observatory was built by Roskosmos in the interests of the Russian Academy of Sciences represented by its Space Research Institute (IKI) in the framework of the Russian Federal Space Program, with the participation of the Deutsches Zentrum f\"ur Luft- und Raumfahrt (DLR). The SRG/eROSITA X-ray telescope was built by a consortium of German Institutes led by MPE, and supported by DLR. The SRG spacecraft was designed, built, launched, and is operated by the Lavochkin Association and its subcontractors. The science data are downlinked via the Deep Space Network Antennae in Bear Lakes, Ussurijsk, and Baykonur, funded by Roskosmos. The eROSITA data used in this work were processed using the eSASS software system developed by the German eROSITA consortium and proprietary data reduction and analysis software developed by the Russian eROSITA Consortium.

This work was supported by the Australian government through the Australian Research Council's Discovery Projects funding scheme (DP200102471).

This work includes data products from the Near-Earth Object Wide-field Infrared Survey Explorer (NEOWISE), which is a project of the Jet Propulsion Laboratory/California Institute of Technology. NEOWISE is funded by the National Aeronautics and Space Administration.

The National Radio Astronomy Observatory is a facility of the National Science Foundation operated under cooperative agreement by Associated Universities, Inc.

Based on observations made with the Nordic Optical Telescope, owned in collaboration by the University of Turku and Aarhus University, and operated jointly by Aarhus University, the University of Turku and the University of Oslo, representing Denmark, Finland and Norway, the University of Iceland and Stockholm University at the Observatorio del Roque de los Muchachos, La Palma, Spain, of the Instituto de Astrofisica de Canarias.
 
\section{Author Contributions}
S.R. first identified \tywin{} as a candidate neutrino source, performed the SED and the dust echo analysis and was the primary author of the manuscript. A.F., M.K., R.St. and S.v.V. contributed to the manuscript, the data analysis and the source modeling. A.F. and M.K. have initiated the ZTF neutrino follow-up program. C.L., K.M. and W.W. performed the neutrino production analysis. J.C.A.M.-J. contributed the VLA observations. M.Gi. performed the SRG/eROSITA observations and data analysis. S.B. and S.Ga. analyzed the Fermi data. S.R. analyzed the Swift-XRT data. S.A., K.D. and S.R. performed and analyzed the P200 IR observations. S.Ge. and S.S. requested and reduced the Swift-UVOT data. V.S.P. contributed the black hole mass estimate. E.Z. contributed to the neutrino event rate calculation. M.G. analyzed the archival radio data. S.S.K and B.T.Z. performed parts of the neutrino production analysis. V.K. and N.L.S. contributed PTF forced photometry. C.B., T.S. and J.S. ruled out other candidates in the follow-up. T.A., M.W.C., M.M.K. and L.P.S. enabled ZTF ToO observations. J.Ne., S.R. and R.St. developed the analysis pipeline. V.B., J.No. and J.v.S. developed AMPEL, and contributed to the ToO analysis infrastructure. E.C.B., S.B.C., R.D., M.J.G., R.R.L., B.R., M.R. and R.W. contributed to the implementation of ZTF. P.M. and R.Su. contributed to the implementation of SRG/eROSITA. S.F., S.Ge., A.G.-Y., A.K.H.K, E.K. and D.A.P. contributed comments/to discussions. All authors reviewed the contents of the manuscript.
\newpage
\section{Supplemental Material}
\label{sec:appendix}
The data and code used for the analysis presented in this paper can be accessed at \url{https://github.com/simeonreusch/at2019fdr} \cite{tywincode}. \texttt{nuztf}, the multimessenger-pipeline used to identify \tywin{} as potential source, can be found here: \url{https://github.com/desy-multimessenger/nuztf}. Both make use of \texttt{AMPEL} \cite{ampel} and \texttt{ztfquery} \cite{ztfquery}.

\paragraph{\textbf{Follow-up of IC200530A}}
\label{app:ic200530a}
The high-energy neutrino IC200530A was the tenth alert followed up with our ZTF neutrino follow-up program. It had an estimated $59\%$ probability of being of astrophysical origin, based solely on the reconstructed energy of 82.2 TeV and its zenith angle \citep{IC200530A}, with a best-fit position of $\text{RA[J2000]}=255.37^{+2.48}_{-2.57}$ and $\text{Dec[J2000]}=+26.61^{+2.33}_{-2.57}$ at 90\% confidence level. The reported localization amounts to a projected rectangular uncertainty area of $25.38 \,\rm deg^{2}$. During ZTF follow-up observations, 87\% of that area was observed (accounting for chip gaps).

For a full overview of all the neutrino alerts followed up as part of our program, see Table \ref{tab:neutrino_alert_overview}. ANTARES did not report any neutrinos from the same direction \cite{antares}, though the corresponding upper limits reported are not constraining for the neutrino production models introduced in this work.

\tywin{}, at a location of RA[J2000] $=257.2786$ and Dec[J2000] $=+26.8557$, was lying within the 90\% localization region of IC200530A. It was consistent with arising from the nucleus of its host galaxy (SDSSCGB 6856.2), with a mean angular separation to the host position as reported in \textit{Gaia} Data Release 2 \citep{gaia_dr2} of $0.03 \pm 0.15$ arcsec. The angular separation from the neutrino best fit position was $1.72$ deg. The event had a spectroscopic redshift of $z=0.267$, which implies a luminosity distance $D_{\rm L} \approx$ 1360 Mpc, assuming a flat cosmology with $\Omega_\Lambda=0.7$ and $H_0=70$ km s$^{-1}$ Mpc$^{-1}$.

\paragraph{\textbf{Gamma ray limits}}
\label{app:gamma ray}
Details on the \textit{Fermi}-LAT analysis and the upper limits can be found in \cite{lancel}. The High Altitude Water Cherenkov Experiment (HAWC) reported that no significant detection was found at the time of neutrino arrival \cite{hawc}. A non-detection was also reported by the International Gamma-Ray Astrophysics Laboratory (INTEGRAL) \cite{integral}.

\begin{figure*}
\centering
\includegraphics[width=1.0\textwidth]{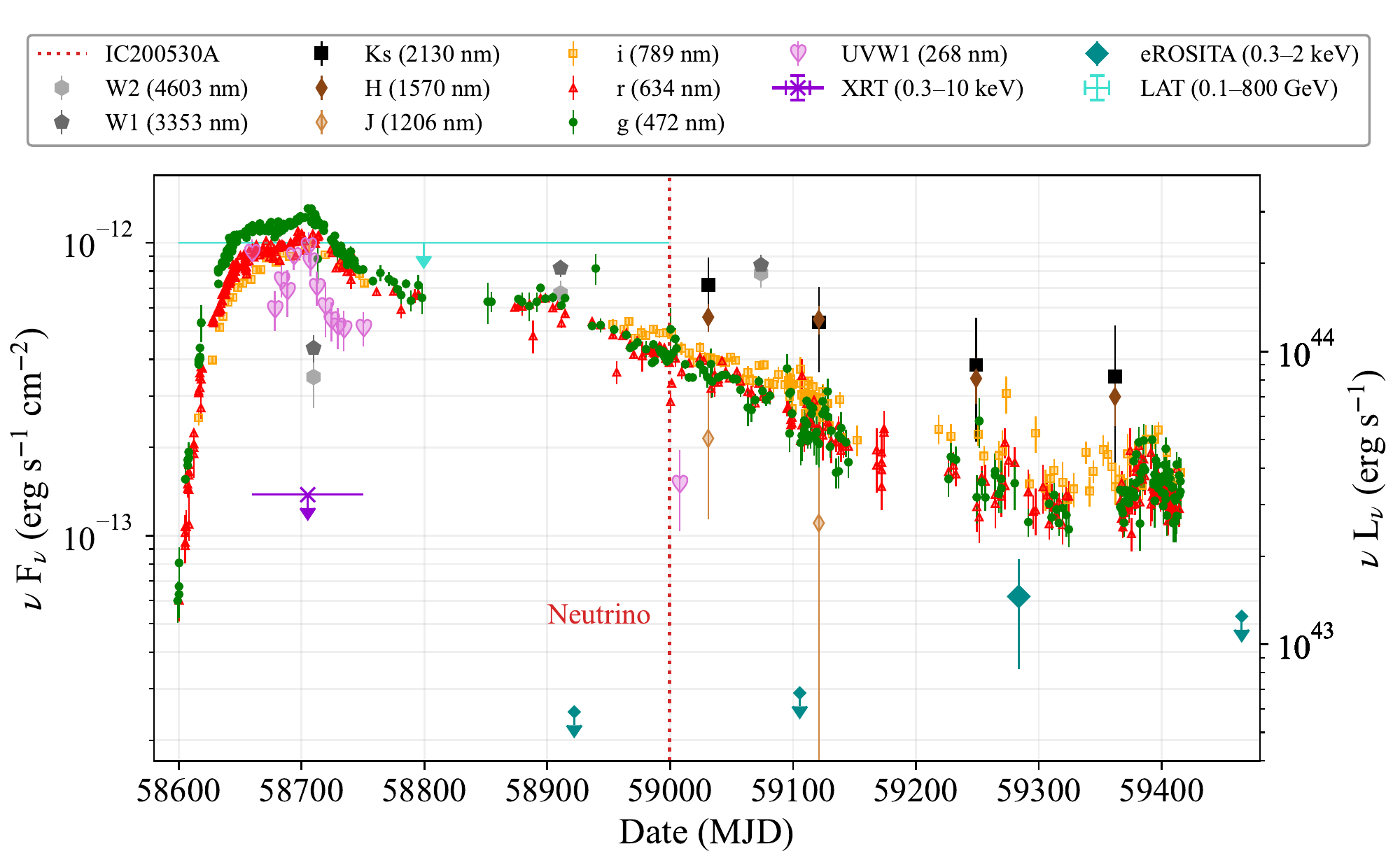}
\caption{Lightcurve of \tywin{}, showing host-subtracted ZTF, \textit{WISE}, P200 and \textit{Swift}-UVOT photometry, non host-subtracted \textit{SRG}/eROSITA photometry, as well as binned \textit{Swift}-XRT and \textit{Fermi}-LAT upper limits (the former has been corrected for absorption). The dotted vertical line indicates the arrival time of the IceCube neutrino IC200530A.}
\label{fig:complete_lightcurve}
\end{figure*}

\paragraph{\textbf{X-ray observations}}
\label{app:X-ray}
In the course of its ongoing all sky survey, the \textit{SRG} observatory \cite{erosita_xrg} visited \tywin{} four times with its 6 month cadence, the first visit having taken place on 2020, March 13--14. The source was detected by eROSITA \cite{erosita} only once, during the third visit on 2021, March 10--11, providing evidence for temporal evolution in the X-ray emission of the source. Constraints on the flux from all three epochs are shown in Fig. \ref{fig:complete_lightcurve} and listed in Table \ref{tab:erosita}. 

The single detection of \tywin{} revealed a very soft thermal spectrum with a best-fit blackbody temperature of $56_{-26}^{+32}\rm\,eV$ (errors are 68\% for one parameter of interest). In the rest frame of the source this corresponds to a temperature of $71_{-33}^{+41}\rm\,eV$ and is among the softest X-ray spectra of all TDEs so far detected by \textit{SRG}/eROSITA \cite{erosita_tde}. The best fit value of the equivalent hydrogen column density ${\rm NH}=1.47_{-1.25}^{+2.80}\times 10^{21}\rm\,cm^{-2}$ is consistent, within the errors, with the Galactic value of ${\rm NH_{\rm Gal}}=0.40\times 10^{21}\rm\,cm^{-2}$ \cite{hi4pi}. As is usually the case for soft sources, there is some degree of degeneracy between the neutral hydrogen column density and the temperature. However, the upper bound on the temperature is still fairly low, $T_{\rm bb}<131$ eV at the 95\% confidence level.

Prior to the detection of IC200530A, \textit{Swift}-XRT had already observed \tywin{} on 14 occasions \citep{sara}. Following the identification of \tywin{} as a candidate neutrino source, an additional prompt ToO observation of the object was requested, which was conducted on 2020, June 7 (2000 second exposure).

To reduce the data and generate a lightcurve, the publicly available \texttt{Swift XRT data products generator} \cite{swift_procedure} was used for the energy range 0.3--10 keV. Further details can be found at \citep{2009MNRAS.397.1177E}. Since the event was not detected in any individual XRT pointings, the 14 observations prior to neutrino arrival were binned (20700 seconds in total) to compute a 3$\sigma$ energy flux upper limit of $1.4 \times 10^{-13} \rm\, erg\,s^{-1}\,cm^{-2}$. The observation after neutrino arrival yielded a 3$\sigma$ energy flux upper limit of $4.7 \times 10^{-12} \rm\, erg\,s^{-1}\,cm^{-2}$. To convert photon counts to energy flux, the \texttt{HEASARC WebPIMMS} tool \cite{heasarc_webpimms} was employed, using the \textit{SRG}/eROSITA blackbody temperature of 56 eV. Absorption was corrected with the best-fit equivalent hydrogen column density from the same measurement (see above).

\begin{table}
\renewcommand{\arraystretch}{1.2}
\centering
\begin{tabular}{c c c  c} 
\setlength{\tabcolsep}{12pt}
\textbf{MJD} & \textbf{Date} & \textbf{Upper limit (95\% CL)} & \textbf{Energy flux}\\
 & & [$\rm erg\,s^{-1}\,cm^{-2}$] & [$\rm erg\,s^{-1}\,cm^{-2}$]\\
\hline
58922 & 2020-03-14 & $ 2.5 \times 10^{-14}$ & -- \\
59105 & 2020-09-13 & $ 2.9 \times 10^{-14}$  & -- \\
59284 & 2021-03-11 & -- & $6.2_{-2.1}^{+2.7}\times 10^{-14}$ \\
59465 & 2021-09-08 & $ 5.3 \times 10^{-14}$ & -- \\
\end{tabular}
\caption{\textit{SRG}/eROSITA detection and upper limits in the 0.3--2.0\,keV band.}
\label{tab:erosita}
\end{table}

\paragraph{\textbf{Optical/UV observations}}
\label{app:Optical/UV}
The ZTF observations were analyzed using dedicated forced photometry, yielding higher precision than `alert' photometry by incorporating knowledge of the transient's position derived from all available images. This was done using the \texttt{ztffps} pipeline \citep{ztffps}, which is built upon \texttt{ztflc} \citep{ztflc}.

To obtain \textit{Swift} measurements, we retrieved the science-ready data from the \textit{Swift} archive (\url{https://www.swift.ac.uk/swift_portal}). We co-added all sky exposures for a given epoch and filter to boost the signal-to-noise ratio using \texttt{uvotimsum} in HEAsoft (\url{https://heasarc.gsfc.nasa.gov/docs/software/heasoft/}, v6.26.1). Afterwards, we measured the brightness of the transient with the \textit{Swift} tool \texttt{uvotsource}. The source aperture had a radius of 3 arcsec, while the background region had a significantly larger radius. The photometry was calibrated with the latest calibration files from September 2020 and converted to the AB system using the methods of \citet{Breeveld2011a}. All measurements were host subtracted using the synthetic host model described in the next section.

\paragraph{\textbf{Infrared observations}}
\label{app:IR}
Four epochs of observations were taken in the J-, H- and Ks-band with the WIRC camera \citep{wirc} mounted on the Palomar P200 telescope in 2020 on July 1 and September 27, as well as in 2021 on February 2 and May 28. The WIRC measurements were reduced using a custom pipeline described in \cite{De2020}. This pipeline performs flat fielding, background subtraction and astrometry (with respect to Gaia DR2) on the dithered images followed by stacking of the individual frames for each filter and epoch. Photometric calibration was performed on the stacked images using stars from the Two Micron All Sky Survey (2MASS) \cite{2006AJ....131.1163S} in the WIRC field to derive a zero point for the stacked images.

The combined host and flare flux was extracted from the stacked images using \texttt{GALFIT} \cite{galfit} to avoid contamination from blending with a close neighboring galaxy. \texttt{photutils} \cite{photutils} was used to derive the Point Spread Function (PSF) for each image. For this purpose, 3--5 isolated stars which were neither dim nor bright were selected by visual inspection of the surrounding area. Based on these, the PSF in the images was fitted. The subtraction quality was verified through visual inspection of the residuals of 4 nearby reference stars from the Sloan Digital Sky Survey (SDSS) \cite{York02}. Following this, S\'ersic profiles were fitted to \tywin{}'s host galaxy and to the neighboring galaxy. Additionally, a point source was fitted.

All parameters except the point source flux were fixed after fitting one epoch (reference epoch). The point source flux was then fit in the other epochs. As the choice of reference epoch had an impact on the flux estimate, both the first and the last epoch were used as such. The difference in the point source flux estimate between both was taken to serve as the systematic uncertainty.

The host galaxy flux was estimated by fitting a galaxy model following the method of \citet{vanVelzen20}. The UV flux was measured with images from the Galaxy Evolution Explorer (\textit{GALEX}) \cite{Martin05}, using the \texttt{gPhoton} software \cite{Million16} with an aperture of 4 arcsec. The optical flux of the host was obtained from the SDSS model magnitudes \cite{Stoughton02}. The baseline \textit{WISE} data points mentioned above were also included, as was an archival measurement from the UKIRT Infrared Deep Sky Survey (UKIDSS) \cite{ukidss}. The \texttt{prospector} software was employed to sample synthetic galaxy models built by Flexible Stellar Population Synthesis (FSPS, \cite{Conroy10, Foreman-Mackey14}). An overview of the values used for constructing the model can be seen in Table \ref{tab:host_model}.

\begin{table*}
\renewcommand{\arraystretch}{1.1}
\centering
\setlength{\tabcolsep}{12pt}
\begin{tabular}{c c c  c  c} 
\textbf{MJD} & \textbf{Date} &\textbf{J-band} & \textbf{H-band} & \textbf{Ks-band} \\
\hline
59031 & 2020-07-01 & $19.06 \pm 0.50$ & $ 17.73 \pm 0.12$ & $17.13 \pm 0.26$\\
59121 & 2020-09-29 & $19.78 \pm 0.98$ & $ 17.76 \pm 0.12$ & $17.45 \pm 0.35$\\
59249 & 2021-02-04 & --               & $ 18.26 \pm 0.19$ & $17.81 \pm 0.48$\\
59362 & 2021-05-28 & -- & $ 18.42 \pm 0.22$ & $17.91 \pm 0.53$\\
\cline{1-4}
 & & \textbf{W1-band} & \textbf{W2-band} & \\
\cline{1-4}
58709  & 2019-08-14 & $17.18 \pm 0.11$ & $17.08 \pm 0.23$ & \\
58910  & 2020-03-02 & $16.49 \pm 0.07$ & $16.37 \pm 0.11$ & \\
59074  & 2020-08-13 & $16.47 \pm 0.09$ & $16.20 \pm 0.11$ & \\
\end{tabular}
\caption{\textit{Top}: NIR AB magnitudes after subtraction of the synthetic host model. Only systematic uncertainties are given (photometric uncertainties should be negligible in comparison, at least in the J- and H-band). The third and fourth J-band epochs had negative flux after host-model subtraction, which are counted as non-detections. \textit{Bottom}: MIR AB magnitudes after subtracting the pre-peak baseline.}
\label{tab:p200_wise}
\end{table*}

\begin{table}
\renewcommand{\arraystretch}{1.3}
\centering
\begin{tabular}{c c } 
\setlength{\tabcolsep}{12pt}
\textbf{Band} & \textbf{AB magnitude} \\
\hline
\textit{GALEX} FUV & $22.32^{+0.07}_{-0.10}$ \\
\textit{GALEX} NUV & $21.52^{+0.06}_{-0.10}$ \\
SDSS u & $20.91^{+0.06}_{-0.07}$ \\
SDSS g & $19.97^{+0.04}_{-0.04}$ \\
SDSS r & $19.00^{+0.02}_{-0.03}$ \\
SDSS i & $18.64^{+0.02}_{-0.03}$  \\
SDSS z & $18.36^{+0.02}_{-0.03}$ \\
UKIRT J & $18.18 $\\
\textit{WISE} W1 & $17.83^{+0.05}_{-0.05}$ \\
\textit{WISE} W2 & $17.78^{+0.03}_{-0.05}$ \\
\end{tabular}
\caption{Archival measurements from \textit{GALEX}, SDSS, UKIRT (the isoMag value was given without error) and \textit{WISE}. These were used to build the synthetic host model.}
\label{tab:host_model}
\end{table}

In addition to the ground-based NIR observations, photometry was extracted from MIR images of the \textit{WISE} \cite{WISE} satellite. For this the W1- and the W2-band were used, centered at 3.4 and 4.6 $\mu$m, respectively. The \textit{WISE} and \textit{NEOWISE} (the new designation after a hibernation period, \cite{NEOWISE}) photometry at the location of \tywin{} suffered from blending with a nearby galaxy, so forced PSF photometry was used \citep{Jones15} on the co-added images \cite{Masci13}. These images were binned in 6 month intervals, aligned to the observing pattern of the satellite. The \textit{WISE} flux in the 13 epochs prior to the optical flare was used to define a baseline, from which difference flux was then computed. We measured a significant flux increase from this baseline. In the W1-band, the root mean square (RMS) variability of the baseline was only 18 $\mu$Jy , which is much smaller than the peak difference flux of 0.9~mJy.

The infrared brightness over time, as detected by P200 and \textit{WISE} after subtracting the host baseline contribution, is shown in Table \ref{tab:p200_wise}.

A selection of IR, optical and UV data are shown in Figure \ref{fig:complete_lightcurve}, alongside upper limits from \textit{Swift}-XRT and \textit{Fermi}-LAT, as well as the measurements from \textit{SRG}/eROSITA.

\paragraph{\textbf{Radio observations}}
We observed \tywin{} with a dedicated Karl G. Jansky Very Large Array (VLA) Director’s Discretionary Time (DDT) program VLA/20A-566 (PI: Stein), and obtained multi-frequency detections at three epochs. The individual observations were taken in 2020 on July 3, September 13 and November 7. The array was in its moderately-extended B configuration for the first two observations, and in the hybrid BnA configuration (with a more extended northern arm) for the final epoch. The first observation was performed in the 2--4 and 8--12\,GHz bands, to which we added the 1--2 and 4--8\,GHz bands in the subsequent epochs. We used \mbox{3C 286} for the delay, bandpass, and flux calibration, and the nearby source J1716+2152 for the complex gain calibration. We followed standard procedures for the external gain calibration, which were performed with the VLA calibration pipeline, using the Common Astronomy Software Application (CASA) v5.6.2. The source was then imaged in each observed band with the CASA task \texttt{tclean}, using Briggs weighting with a robust factor of 1 as a compromise between sensitivity and resolution. The target flux density was measured by fitting a point source in the image plane.

\begin{figure}
\centering
\includegraphics[width=1\columnwidth]{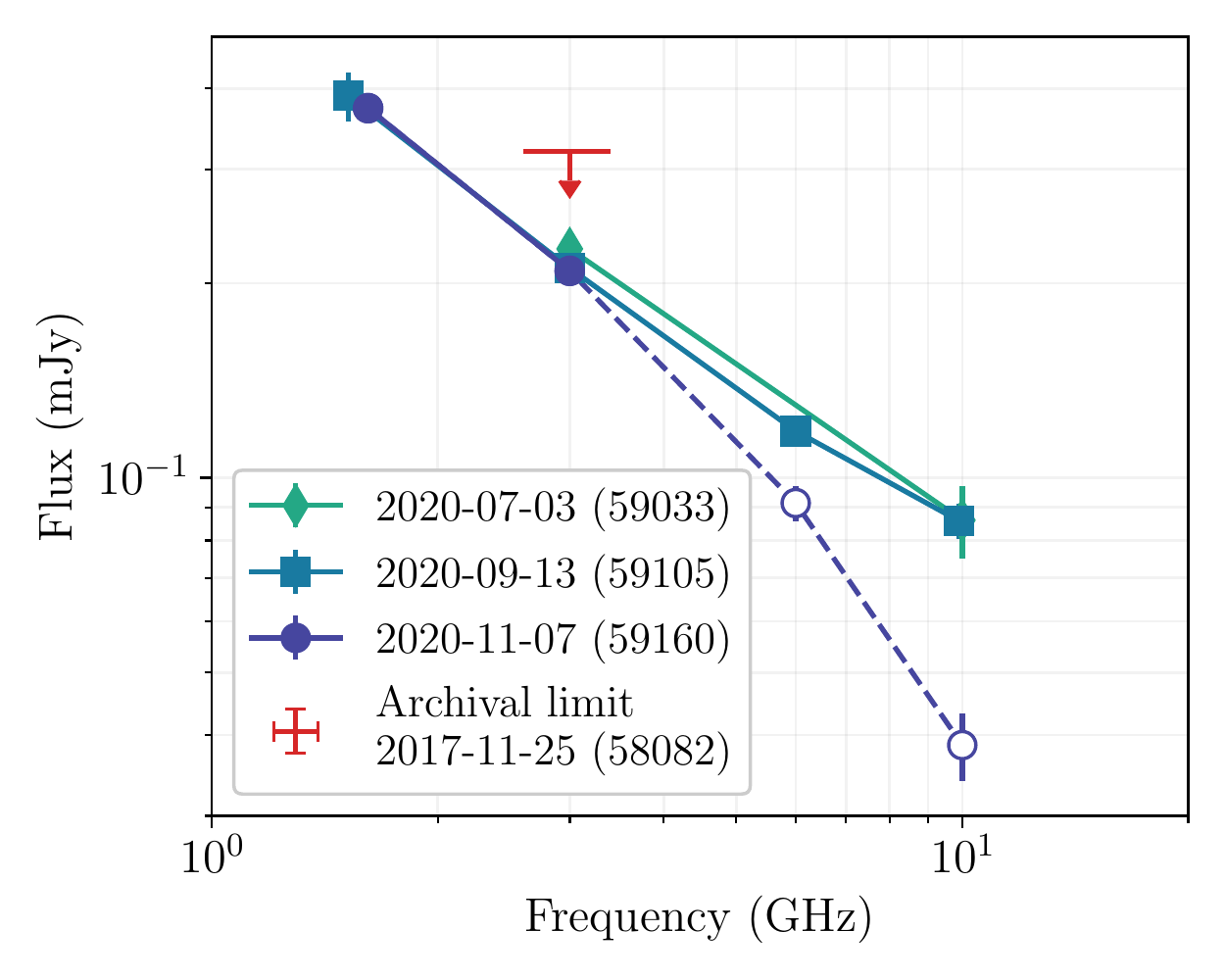}
\caption{Three epochs of VLA measurements of \tywin{}, as well as the VLASS archival upper limit. Note: The apparent spectral steepening in our third epoch (empty circular markers) appears not to be intrinsic to the source, but due to systematic effects.}
\label{fig:radio}
\end{figure}

While consistent flux densities were measured in the 2--4 and 8--12\,GHz bands in the first two observations, the final epoch suggested a possible spectral steepening, with reduced flux densities in the 4--8 and 8--12\,GHz bands (see Fig. \ref{fig:radio}). However, extensive testing revealed that this apparent spectral steepening was not intrinsic to the source. We interpolated the gains derived on the first, third and last of the five 8--12\,GHz scans on J1716+2152 to the remaining two scans, deriving their calibration in the same way as for \bran{} (see \cite{Bran}). The measured calibrator flux densities for those two scans were found to be significantly lower than for the other three. This implied significant atmospheric phase changes between calibrator scans, which reduced the measured flux densities due to decorrelation. Without sufficient flux density in the target field to apply self-calibration, we were unable to correct for this effect, which is worst at the highest frequencies. Therefore, no evidence for intrinsic source variability can be found in the radio band across the 5 months of observations.

Additionally, an archival upper limit was obtained from the Very Large Array Sky Survey (VLASS) \cite{vlass}, which was the only sky survey with adequate sensitivity and angular resolution to probe the emission on the angular and flux density scales relevant for \tywin{}. The quicklook continuum fits image for tile T17t23We was downloaded from the archive (\url{https://archive-new.nrao.edu/vlass/quicklook/}) and dates to 2017, November 25 (2--4\,GHz band). No emission was detected at a $3\sigma$ significance, where $\sigma = 0.11$ mJy/beam is the local RMS noise and the beam is 2.46 arcsec $\times$ 2.28 arcsec (position angle $=-37^\circ$), resulting in an upper limit of $0.32\rm\,mJy$. This result is fully consistent with what was measured in the same band in our dedicated observations. A more recent observation from VLASS epoch 2 was taken very close to the second dedicated observation (2020, September 6), but the local noise was slightly larger and the $3\sigma$ upper limit was less constraining ($0.4\rm\,mJy$).

All our measurements, as well as the archival limit from VLASS are shown in Fig. \ref{fig:radio} and Table \ref{tab:radio}.
 
The observed radio spectra can be described by synchrotron emission from a population of relativistic electrons. These electrons were assumed to be accelerated into a power-law distribution in energy, which leads to a power law in the unattenuated synchrotron spectrum. No break in the power-law spectrum was observed. Therefore the peak frequency, where synchrotron self-absorption sets in, is expected to lie below the lowest observed frequency. The data was analyzed using the models from \cite{2013ApJ...772...78B, Bran}. This allowed to infer energies of a few times $10^{50}$ erg for the electron population, under the assumption of equipartition; a peak flux of 0.5\,mJy at 1\,GHz; and neglecting the impact of baryons. This value provides a lower limit on the non-thermal energy of the system. Given that electrons are typically accelerated with much lower efficiency than protons in astrophysical accelerators \cite{2012A&A...538A..81M}, they were assumed to carry 10\% of the energy carried by relativistic protons ($\epsilon_{e} = 0.1$). Furthermore, the magnetic field was assumed to carry $0.1\%$ of the total energy ($\epsilon_{B} = 10^{-3}$), as indicated by radio observations of TDEs \citep{2018ApJ...854...86E} and supernovae \citep{2013MNRAS.436.1258H}. Under these assumptions, the total energy energy in non-thermal particles was inferred to be $\sim10^{52}$ erg. The inferred size of the emission region, $\sim10^{18}$ cm, and the lack of temporal evolution suggest that the emission was primarily powered by an outflow that was active already prior to the transient event. A second, sub-dominated component could be present. Since the radio observations cover only a period of 5 months, a fast-varying transient signal would be easier to constrain than a slower one. For instance, a relativistic jet viewed face-on would show faster variations and dominate the radio flux if sufficiently energetic, while an off-axis relativistic jet could escape detection due to Doppler deboosting of its radio flux (the latter explanation only works up to the point the jet decelerates and starts to emit isotropically, see e.g. \cite{Generozov17}).

\begin{table}
\centering
\setlength{\tabcolsep}{12pt}
\begin{tabular}{c c  c  c} 
\textbf{MJD} & \textbf{Date} &\textbf{Band} & \textbf{Flux density}\\
& & [GHz] & [$\mu$ Jansky] \\
\hline
58032 & 2017-11-25 & 3.00 & $\leq 320 $ \\ 
59033 & 2020-07-03 & 3.00 & $ 226 \pm 13$ \\
59033 & 2020-07-03 & 10.00 & $ 86 \pm 11$ \\
59105 & 2020-09-13 & 1.52 & $ 390 \pm 34$ \\
59105 & 2020-09-13 & 3.00 & $ 211 \pm 10$ \\
59105 & 2020-09-13 & 6.00 & $ 118 \pm 6$ \\
59105 & 2020-09-13 & 10.00 & $ 86 \pm 5$ \\
59160 & 2020-11-07 & 1.62 & $ 373 \pm 20$ \\
59160 & 2020-11-07 & 3.00 & $ 209 \pm 9$ \\
59160 & 2020-11-07 & 6.00 & $ 91 \pm 6$ \\
59160 & 2020-11-07 & 10.00 & $ 39 \pm 5$ \\
\end{tabular}
\caption{VLA measurements and the VLASS archival limit (first row).}
\label{tab:radio}
\end{table}
 
\paragraph{\textbf{Double-blackbody fit}}
\label{app:double_bb}
The SED was fit using the \texttt{lmfit} Python package \citep{lmfit} with both a broken and a non-broken intrinsic power law, as well as a single and a double unmodified blackbody spectrum. The data are well described by a double-blackbody model, comprised of an optical/UV `blue' blackbody and an infrared `red' blackbody. Fitting is done for three epochs during which infrared, optical and UV data is available. These epochs in MJD are: 1) 58700--58720, 2) 59006--59130 and 3) 59220--59271. To account for host extinction, we employed \texttt{extinction} \cite{extinctionpackage} using the Calzetti attenuation law \cite{calzetti}. The extinction parameter $A_V$ was left free in epoch 1 and was fixed at the epoch 1 best-fit value ($A_V = 0.45^{+0.14}_{-0.14}$ mag for $R_V = 3.1$) for epochs 2 and 3. We chose epoch 1 to determine the extinction parameter because the observations of the optical/UV are most precise there.

The other fit parameters were the blackbody temperature and the the blackbody radius, resulting in a total of 6 fit parameters in epoch 1, and 4 in epochs 2 and 3. The model was fitted to the data in the three epochs using the Levenberg-Marquardt minimization algorithm. 68\% confidence levels for all parameters were estimated by determining $\Delta \chi^{2}=1$ relative to the best-fit $\chi^{2}$, stepping through each parameter while leaving the other parameters free during minimization.

The best fit blackbody values are shown in Table \ref{tab:double_bb}; the respective SEDs for the three epochs are shown in the top panels of Fig. 1 in the main text.

\begin{table}
\renewcommand{\arraystretch}{1.3}
\centering
\begin{tabular}{c c  r  c  c} 
 \textbf{Epoch} &  \textbf{Band}  & \textbf{Temp.} & \textbf{Radius} & \textbf{Luminosity} \\
 & & [K] & [cm]&  [erg s$^{-1}$] \\
\hline
\textbf{1} & O+UV & $ 13526^{+569}_{-574}$ & $ 7.8^{+0.4}_{-0.4} \times 10^{15}$ & $ 1.4^{+0.1}_{-0.1} \times 10^{45}$\\
 & IR  & $1505^{+421}_{-313}$ & $ 2.2^{+1.6}_{-1.4} \times 10^{17}$ & $1.7^{+2.2}_{-1.0} \times 10^{44}$\\
\hline
\textbf{2} & O+UV & $ 11731^{+663}_{-683}$ & $ 4.9^{+0.4}_{-0.4} \times 10^{15}$ & $ 3.3^{+0.3}_{-0.4} \times 10^{44}$\\
 & IR  & $1762^{+121}_{-124}$ & $ 2.5^{+0.2}_{-0.2} \times 10^{17}$ & $4.3^{+0.5}_{-0.7} \times 10^{44}$\\
\hline
\textbf{3} & O+UV & $ 10230^{+2373}_{-1645}$ & $ 4.3^{+3.3}_{-1.0} \times 10^{15}$ & $ 1.5^{+1.2}_{-0.4} \times 10^{44}$\\
 & IR  & $2237^{+402}_{-462}$ & $ 1.0^{+0.6}_{-0.4} \times 10^{17}$ & $1.9^{+1.4}_{-0.5} \times 10^{44}$\\
\end{tabular}
\caption{Blackbody fit values for three epochs (1--3), where `O+UV' denotes the blue blackbody in the optical/UV and `IR' denotes the red infrared blackbody. The luminosity is given dereddened and in the source frame and the uncertainties are at 68\% confidence level. Note that the O+UV temperature and radius (and therefore the luminosity) in the third epoch are not well constrained, as there are no late-time UV measurements available. The same holds true for the infrared blackbody in the first and the last epoch, as only 2 data points are available in the infrared.}
\label{tab:double_bb}
\end{table}

\paragraph{\textbf{Dust echo model and energy output}}
Following the procedure in \cite{dustecho}, the optical lightcurve was convolved with a rectangular function of width $2 \times \Delta t_c$, where $\Delta t_c$ is the light travel time from the transient to the surrounding dust region. A best-fit value of $\Delta t_c = 193$ days was obtained, which translates to a distance to the dust region of $5 \times 10^{17}\rm\,cm$ ($0.16\rm\,pc$). To obtain a measure for the peak luminosity of the transient, the blue blackbody fit at peak was used. The resulting unobscured peak luminosity from the blue blackbody was $L_{\rm peak} = 1.4^{+0.1}_{-0.1} \times 10^{45}\rm\,erg\,s^{-1}$. To check this value for consistency, the peak optical/UV luminosity was also calculated following \cite{dustecho}. This method relies only on the blackbody temperature (normalized to $1850\rm\,K$) and the radius of the dust region, normalized to $0.1\rm\,pc$ (assuming a dust grain size of $0.1\,\mu\rm m$): 
\begin{equation*}
L_{\rm peak} = 5 \times 10^{44} R^2_{0.1} T^{5.8}_{1850} \rm\,erg\,s^{-1}
\end{equation*}
This yielded a value of $L_{\rm peak} = 1.3\times10^{45}\rm\,erg\,s^{-1}$, which is in good agreement with the value derived from the blackbody fit described in the section above. 

As the temperature of the optical/UV blackbody decreases only slightly and stays roughly centered on the g-band, the total energy radiated in the optical/UV can be approximated by scaling the peak luminosity with the g-band lightcurve and integrating over time. The resulting energy output was $E_{\rm O+UV} = 3.4 \times 10^{52}\rm\,erg$.

The peak dust echo luminosity was inferred from the IR blackbody fit to be $L_{\rm dust} = 4.3 \times 10^{44}$ erg s$^{-1}$. Similarly to the optical/UV, we time-integrated the fitted dust echo lightcurve, normalized to the peak dust echo luminosity from epoch 2, to obtain the total bolometric energy of the dust echo: $E_{\rm dust} = 1.1 \times 10^{52}$ erg. The ratio of the two time-integrated energies resulted in an unusually high covering factor of $1/3$, while the TDE dust echoes in quiescent galaxies typically show covering factors of around 1\% \cite{dustecho}. 


\paragraph{\textbf{Black hole mass}}
\label{app:bh_mass}
The commonly adopted method to derive the mass of the central SMBH ($M_{\rm BH}$) of an AGN is using its optical spectrum and is called the single-epoch virial technique (see e.g. \cite{VP06}). The NOT spectrum used for this was taken on 2020, April 30 with the ALFOSC camera. There are later spectra, but these do not contain the H$\alpha$ emission line. The spectrum was reduced in a standard way, consisting of wavelength calibration through an arc lamp and flux calibration using a spectrophotometric standard star.

\begin{figure*}
  \centering
    \includegraphics[width=0.8\textwidth]{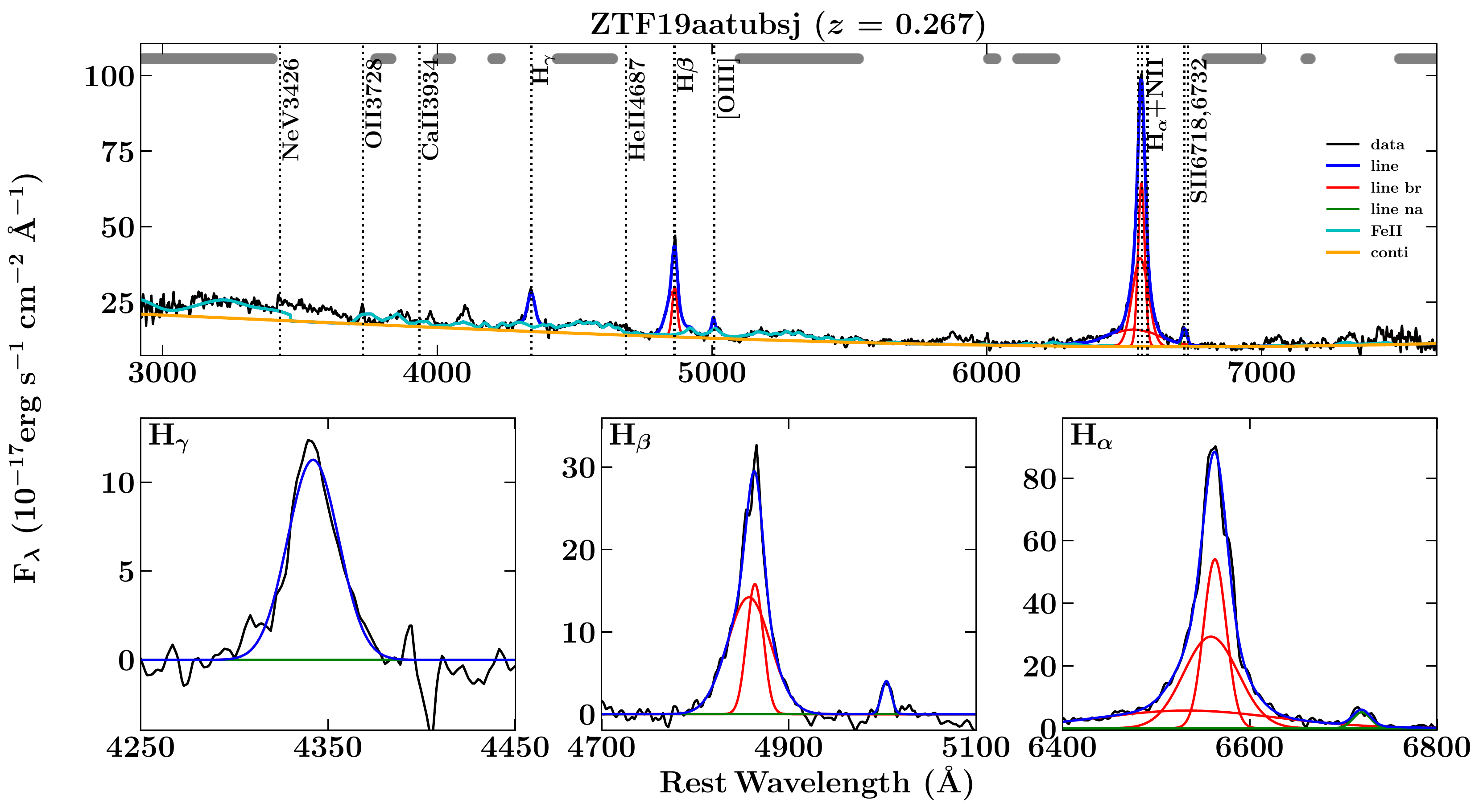}
    \caption{The top plot shows the NOT spectrum of \tywin{} analyzed with {\tt PyQSOFit}; different line components are labeled. The bottom panels show zoomed versions of the fitted line complexes.}
    \label{fig:spec_fit}
\end{figure*}

The publicly available multi-component spectral fitting tool {\tt PyQSOFit} \cite{Guo18} was used for spectral analysis. We brought the spectrum to the lab frame and dereddened it using the Milky Way extinction law of \cite{F99} with $R_{\rm V}=3.1$ and dust map data adopted from \cite{dust} -- note that the extinction law is not the same as the one used in the blackbody fits, but the difference is small. The continuum was modeled using line-free regions as a combination of a third degree polynomial and an optical Fe II template adopted from \cite{BG92}. It was subtracted from the spectrum to obtain the line spectrum only. The H$\alpha$ and H$\beta$ line complexes were then fitted separately, while simultaneously fitting the emission lines in each complex. In the wavelength range of [6400,6800] \AA, the H$\alpha$ emission line was modeled with three Gaussian functions, two for the broad components and one for the narrow. [N {\sc ii}] $\lambda\lambda$6548,6584, and [S {\sc ii}] $\lambda\lambda$6717,6731 were also fitted with a single Gaussian. The H$\beta$ emission line was fitted in the wavelength range [4700,5100] \AA, similar to the H$\alpha$ line. Finally, the [O~{\sc iii}] $\lambda\lambda$4959,5007 doublet was modeled with two Gaussian functions. The estimated uncertainties are statistical only and do not include the intrinsic scatter (0.3$-$0.4 dex) associated with the virial approach (see e.g. \cite{VP06,Shen11}). Further details about the fitting technique and {\tt PyQSOFit} can be found in \cite{Guo19} and \cite{2019ApJS..241...34S}. The fitted spectrum is shown in Figure~\ref{fig:spec_fit}.

The measured full-width-at-half-maximum (FWHM) of the H$\beta$ line was $1914 \pm 282$ km s$^{-1}$ and the continuum luminosity at 5100 \AA~was $\lambda L_{\lambda}=44.16\pm0.01$ (log-scale, erg s$^{-1}$). Broadening due to limited detector resolution was considered to be small. From the estimated FWHM and $\lambda L_{\lambda}$, the mass of the central black hole was computed by adopting the following empirical relation \cite{Shen11}:

\begin{equation*}\label{eqn:virial_estimator}
\scriptstyle
\log \left( \frac{M_{\rm BH,vir}}{M_\odot} \right) = a + b \log \left( \frac{\lambda L_{\lambda}}{10^{44}\,{\rm
erg/s}} \right) + 2 \log \left( \frac{\rm FWHM}{\rm km/s} \right)
\end{equation*}

The coefficients $a$ and $b$ were taken as 0.91 and 0.5, adopted from \cite{VP06}. This resulted in $M_{\rm BH} = 10^{7.55 \pm 0.13} M_\odot$. \cite{Shen11} also give the following empirical relation to estimate $M_{\rm BH}$ from the H$\alpha$ line FWHM and luminosity:

\begin{equation*}
\scriptstyle
\log \left( \frac{M_{\rm BH,vir}}{M_\odot} \right)_{\rm H\alpha} = ~  0.379 + 0.43 \log \left( \frac{L_{\rm H\alpha}}{10^{42} \rm erg/s}\right) + 2.1 \log \left( \frac{\rm FWHM_{H\alpha}}{\rm km/s} \right)
\end{equation*}

Supplying the derived H$_{\alpha}$ FWHM ($1753 \pm246 \rm\,km\,s^{-1}$) and luminosity (43.05$\pm$0.06, log-scale, erg s$^{-1}$) in the above equation, $M_\text{BH}$ was computed as $10^{7.64 \pm 0.13} M_\odot$, consistent with the previous estimate.

These estimates are consistent with the masses derived in the NLSy1 ZTF study \cite{sara}, which employed two different methods. The virial approach (also used above) yielded an estimate of $10^{7.1} M_\odot$, while an approach using the host galaxy luminosity according to \cite{mclure} resulted in $10^{7.8} M_\odot$. In this paper an Eddington ratio of $\frac{L}{L_\text{Edd}} = 0.24-1.2$ was derived. The different result for the virial approach can be explained by the fact that \cite{sara} fit a single Lorentzian profile while here a multi-Gaussian approach is used. Also, the spectrum employed for the fit presented here was more recent, presumably containing less transient emission.

\paragraph{\textbf{Chance coincidence}}
To compute the chance coincidence of finding an event like \tywin{} in association with a high-energy neutrino, we obtained the full set of nuclear ZTF transient flares as selected by \mbox{AMPEL} \cite{ampel, ztf_alertsystem} from the ZTF alert stream \cite{ztf_dataprocessing}. At least 10 detections in both g- and r-band were required, as was a weighted host--flare offset $\leq 0.5$ arcsec and a majority of the data points having positive flux after subtracting the reference image. The dataset was further restricted to transients being first detected after 2018 January 1 and peaking before July 2020. 3172 flares survived these cuts.

We further required the nuclear transients not to be classified as (variable) star or bogus object (e.g. subtraction artifacts); additionally, flares were rejected when their rise (fade) $e$-folding time (see \cite{winterishere} for details) was smaller than the uncertainty on this value; these two cuts left 1628 candidates. As we were only interested in events of brightness comparable to \tywin{}, we required that the peak apparent magnitude (see \cite{lancel} for details) $\leq$ the peak magnitude of \tywin{}; this left 157 events. Furthermore, the rise (fade) $e$-folding time was required to be in the [15, 80] ([30, 500]) day interval to select for (candidate) TDEs, resulting in 25 transients.

Nuclear transients which have been spectroscopically classified as supernovae were excluded from the remaining 25 events. Furthermore, we visually excluded transients showing only short-timescale AGN variability, displaying no consistent color or color evolution and events which were not sufficiently smooth post peak. This procedure left a final sample of 12 transients.

To calculate the effective source density of these 12 events, their lifetime was conservatively estimated at 1 year per event. The ZTF survey footprint is $28000\rm\,deg^{2}$ (excluding sources with galactic latitude $|b|<7$) \cite{Bran}. From the time range of the sample (2.5 years) and the ZTF footprint, the effective source density was computed as $1.7\times10^{-4}\rm\,deg^{-2}$. This is the density of sources per $\rm\,deg^{2}$ of sky in the survey footprint at any given time. Through multiplying the effective source density by the combined IceCube alert footprint of $154.3\rm\,deg^{2}$, an expectation value for the number of neutrinos can be calculated. Employing a Poisson distribution, the chance coincidence of finding 2 sources by chance was calculated as $p=3.4 \times 10^{-4}$ ($p=2.6 \times 10^{-2}$ for one source only).

\begin{table*}
\centering
\small
	\begin{tabular}{l r r r r r c} 
		\textbf{Event} & \textbf{R.A. (J2000)} & \textbf{Dec (J2000)} & \textbf{90\% area} & \textbf{ZTF obs} &~ \textbf{Signal-}& \textbf{Reference}\\
		& \textbf{(deg)}&\textbf{(deg)}& \textbf{(deg$^{2}$)}& \textbf{(deg$^{2}$)} & \textbf{ness} &\\
		\hline
		IC190503A & $120.28$ & $6.35$ & 1.94 & 1.37 & 36\%&\cite{IC190503A1, IC190503A2}\\
	    IC190619A & $343.26$ & $10.73$ & 27.21 & 21.57 & 55\%&\cite{IC190619A1, IC190619A2}\\
		IC190730A & $225.79$ & $10.47$ & 5.41 & 4.52 & 67\%&\cite{IC190730A1, IC190730A2}\\
		IC190922B & $5.76$ & $-1.57$ & 4.48 & 4.09 & 51\%&\cite{IC190922B1, IC190922B2, IC190922B3}\\
	    IC191001A & $314.08$ & $12.94$ & 25.53 & 23.06 & 59\%& \cite{IC191001A1, IC191001A2, IC191001A3}\\
	    IC200107A & $148.18$ & $35.46$ & 7.62 & 6.28 & -- &\cite{IC200107A1, IC200107A2}\\
    	IC200109A & $164.49$ & $11.87$ & 22.52 & 22.36 & 77\%&\cite{IC200109A1, IC200109A2}\\
        IC200117A & $116.24$ & $29.14$ & 2.86 &  2.66 & 38\%&\cite{IC200117A1, IC200117A2, IC200117A3}\\
        IC200512A & $295.18$ & $15.79$ & 9.77 &  9.26 & 32\%&\cite{IC200512A1, IC200512A2}\\
        \textbf{IC200530A} & \textbf{255.37} & \textbf{26.61} & \textbf{25.38} &  \textbf{22.05} & \textbf{59\%}&\cite{IC200530A1, IC200530A2, IC200530A3, IC200530A4}\\
        IC200620A & $162.11$ & $11.95$ & 1.73 & 1.24 & 32\%&\cite{IC200620A1, IC200620A2}\\
        IC200916A & $109.78$ & $14.36$ & 4.22 & 3.61 & 32\%&\cite{IC200916A1, IC200916A2, IC200916A3}\\
        IC200926A & $96.46$ & $-4.33$ & 1.75 & 1.29 & 44\%&\cite{IC200926A1, IC200926A2}\\
        IC200929A & $29.53$ & $3.47$ & 1.12 & 0.87 & 47\%&\cite{IC200929A1, IC200929A2}\\
        IC201007A & $265.17$ & $5.34$ & 0.57 & 0.55 & 88\%&\cite{IC201007A1, IC201007A2}\\
        IC201021A & $260.82$ & $14.55$ & 6.89 & 6.30 & 30\%&\cite{IC201021A1, IC201021A2}\\
        IC201130A & 30.54 & $-12.1$ & 5.44 & 4.51 & 15\%&\cite{IC201130A1, IC201130A2}\\
        IC201209A & $6.86$ & $-9.25$ & 4.71 & 3.20 & 19\%&\cite{IC201209A1, IC201209A2}\\
        IC201222A & $206.37$ & $13.44$ & 1.54 & 1.40 & 53\%&\cite{IC201222A1, IC201222A2}\\
        IC210210A & $206.06$ & $4.78$ & 2.76 & 2.05 & 65\%&\cite{IC210210A1, IC210210A2}\\
        IC210510A & $268.42$ & $3.81$ & 4.04 & 3.67 & 28\%&\cite{IC210510A1, IC210510A2}\\
        IC210629A & $340.75$ & $12.94$ & 5.99 & 4.59 & 35\%&\cite{IC210629A1, IC210629A2}\\
        IC210811A & $270.79$ & $25.28$ & 3.17 & 2.66 & 66\%&\cite{IC210811A1, IC210811A2}\\
        IC210922A & $60.73$ & $25.28$ & -4.18 & 1.16 & 93\%&\cite{IC210922A1, IC210922A2}\\
	\end{tabular}
	\caption{Summary of the 24 neutrino alerts followed up by ZTF until September 2021, with IC200530A highlighted. The 90\% area column indicates the rectangular uncertainty region of sky as reported by IceCube. The ZTF obs column indicates the area observed at least twice by ZTF, within the reported 90\% localization (accounting for chip gaps). \textit{Signalness} estimates the probability that the neutrino is of astrophysical origin, rather than caused by atmospheric background. The total followed-up area (corrected for chip-gaps) is 154.33 deg$^{2}$. }
	\label{tab:neutrino_alert_overview}
\end{table*}

\paragraph{\textbf{Disk-corona model}}
Accretion disks and their coronae have been proposed as neutrino production sites (see \cite{murase_bran}). 
The plasma is expected to be collisional for standard disks and slim disks. In particular, for super-Eddington disks, the proton-electron relaxation time is shorter than the infall time, leading to $kT_e\approx kT_p$. So sub-PeV neutrino production inside accretion disks including those from super-Eddington magnetically arrested disks~\cite{2019ApJ...886..114H} are highly suspicious for this TDE. On the other hand, it has been shown that the plasma can be collisionless for highly magnetized regions such as disk-coronae and radiatively inefficient accretion flows \cite{murase_seyfert, 2015ApJ...806..159K}. Not only magnetic reconnections but also the Fermi acceleration mechanism may operate if the plasma density is not too large to be collisionless. This work focuses on neutrino emission from the coronae. 

Following \cite{murase_bran}, we calculated the neutrino emission from coronae above the TDE disk. The size of the coronal region was assumed to be $R=30\,R_S$, where $R_S$ is the Schwarzschild radius, the alpha viscosity was set to $\alpha=0.1$, and the plasma $\beta$ was assumed to be $\beta=0.2$. Spectral energy distributions were determined by the optical and ultraviolet luminosity, $\nu F_\nu=10^{-12}~{\rm erg}~{\rm cm}^{-2}~{\rm s}^{-1}$. This is consistent with a bolometric luminosity of $L\sim{10}^{45}~{\rm erg}~{\rm s}^{-1}$, which leads to the Eddington ratio $\lambda_{\rm Edd}\equiv L_{\rm bol}/L_{\rm Edd}\sim 0.3$ (the accretion is near the Eddington limit, and super-Eddington accretion is often characterized by $\lambda_{\rm Edd}\gtrsim1$~\cite{McKinney+15,Jiang+19,Takeo+19}).
The corresponding Comptonized X-ray luminosity was around a few ${10}^{43}~{\rm erg}~{\rm s}^{-1}$, which can be reprocessed by the TDE debris. 

Neutrinos and gamma rays can be efficiently produced by both $p\gamma$ and $pp$ interactions \cite{murase_seyfert}. Here the \texttt{SOPHIA} Monte Carlo code \cite{2000CoPhC.124..290M} was used to simulate neutrino spectra from $p\gamma$ interactions. For non-thermal protons, stochastic acceleration was assumed, although magnetic reconnections may also be considered. The acceleration time is given by $t_{\rm acc}\approx \eta_B {(c/V_A)}^{2}(H/c) {(r_{L}/H)}^{2-q}$, where $V_A$ is the Alfv\'en velocity, $H$ is the scale height, and $r_L$ is the Larmor radius. We used $\eta_B=10$ and $q=5/3$.

The cosmic ray luminosity was normalized by the cosmic ray pressure, for which $P_{\rm CR}=0.5 n_pk_B T_{p}$ was used, where $n_p$ is the proton density and the proton temperature $T_p$ was assumed to be the virial temperature. This gives a reasonable upper limit in accretion flow models. The expected number of muon neutrinos under the effective area of the neutrino alert channel is $\sim0.007$. The peak neutrino energy depends on $\beta$, and the observed energy of $\sim80$~TeV can be achieved even for $\beta\sim0.5-1$. Note that in general the cosmic ray luminosity is expected to be less than the optical/UV accretion disk luminosity, $L_{\rm CR}\lesssim L$ because of $\eta_{\rm CR}\lesssim \eta_{\rm rad}$. 

In this model, there are two explanations for the time delay between optical peak and neutrino detection~\cite{murase_bran}. The accretion at the earliest phase may be super-Eddington, but it will enter the sub-Eddington regime as the accretion rate decreases. Recent magnetohydrodynamic simulations have shown that high-temperature coronal regions form above the accretion disk \cite{2014ApJ...780...46I, 2014ApJ...784..169J}. The neutrino observation time can be consistent with the formation time of the collisionless corona. Alternatively, the corona may form even in the super-Eddington phase \cite{2016ApJ...826...23T}, with the neutrino luminosity being initially saturated and tracing the accretion history at late times.

\paragraph{\textbf{Sub-relativistic wind model}}
Non-relativistic outflows provide natural sites for high-energy neutrino and gamma-ray production. It has been believed that a TDE event is accompanied by the formation of an accretion disk, launching disk-driven winds with a velocity of $V_w\sim0.1 c$. The observed optical emission could be reprocessed emission of shocked dissipation by the wind \cite{Strubbe:2009qs,Miller:2015jda,Metzger:2015sea} and the launch of radiation-driven winds has been supported by numerical simulations (e.g. \cite{Jiang+19}).

Self-interactions between tidal streams and/or collisions between the wind and TDE debris would lead to a strong shock, in which particles (both ions and electrons) can be accelerated. If the magnetic field energy density is $U_B\equiv\epsilon_B L_w/(4\pi R_w^2 V_w)$, then $\epsilon_B\sim0.03$, a wind kinetic luminosity of $L_w\sim10^{45.5}~{\rm erg}~{\rm s}^{-1}$ and a termination shock radius of $R_w\sim10^{16}$~cm lead to $B_w\sim30$~G and a maximum proton energy of 25~PeV in the presence of $p\gamma$ energy losses. Then, cosmic rays leaving the wind interact with the TDE debris via inelastic $pp$ collisions. For a TDE event of a main-sequence star with $\sim 1~M_\odot$, the effective optical depth for neutrino production is nearly unity at the assumed dissipation radius, $R_{\rm diss}=R_w=10^{16}$~cm, in which the system is approximately calorimetric for neutrino and gamma-ray production~\cite{murase_bran}. We use the optical/UV luminosity $L=1.4\times{10}^{45}~{\rm erg}~{\rm s}^{-1}$ from the blackbody fit and $L_X=0.03\, L$, although in this model the results are not sensitive to X-rays. The magnetic field is set to $B_w=30$~G. 

The cosmic ray luminosity was normalized by the baryon loading factor, $\xi_{\rm CR}=L_{\rm CR}/L$. For $\xi_{\rm CR}=3$, the expected number of neutrino events was $\sim0.002$ under the effective area of the neutrino alert channel. The resulting neutrino spectrum consists of $pp$ and $p\gamma$ components, and there is a dip around $\sim100$~TeV caused by the Bethe-Heitler pair production due to optical and ultraviolet photons. In this model, the neutrino light curve is expected to trace the kinetic luminosity as long as the system is calorimetric~\cite{murase_bran}.

\paragraph{\textbf{Jet model}}
\label{app:TDE_concordance}

Following \citet{Winter:2020ptf}, we considered a relativistic jet, where the beamed neutrino emission allows to naturally meet the energy requirement. In the absence of X-ray signatures before and near the neutrino detection time, the model has been based on a TDE unified model~\cite{Dai:2018jbr}. 

The interpretation of TDE X-ray signals over a wide range of SMBH masses has been the subject of a recent series of papers \cite{Mummery:2021otz,Mummery:2021qrc,Mummery:2021nqy,Mummery:2021rgp}. In short, for higher SMBH masses, lower X-ray temperatures (possibly including a non-thermal tail) are expected to shift the peak of the thermal spectrum coming from the disk out of the \textit{Swift} or \textit{SRG}/eROSITA energy ranges. Following \citet{Mummery:2021nqy} for the temperature and luminosity dependence on these quantities, we extrapolated from the values for \bran{} (where $M \simeq 10^6~M_\odot$ is assumed) to the corresponding values for $M \sim 10^7~M_\odot$, obtaining an X-ray temperature $T_X \simeq 34 \, \mathrm{eV}$ and an X-ray peak luminosity $L_X \simeq 1.7 \times 10^{43} \, $erg s$^{-1}$ in the \textit{Swift} or \textit{SRG}/eROSITA  energy windows ($L_{\mathrm{FUV}} \simeq 7 \times 10^{44}$ erg s$^{-1}$ at the BB peak, bolometrically corrected over the full energy range). The disk X-ray luminosity was assumed to scale with the BB luminosity as well, following its lightcurve.
The X-ray peak luminosity assumption is accidentally consistent with the late-time X-ray detection luminosity within uncertainties ($L_X \simeq 1.4 \times 10^{43} \, $erg s$^{-1}$ in the \textit{SRG}/eROSITA range at MJD 59284), and it is roughly consistent with the earlier limits if a moderate amount of time evolution or X-ray obscuration is assumed (which is required independent of the model). For the X-ray measurements, see Fig.~\ref{fig:complete_lightcurve} and Table \ref{tab:erosita}.

Note that it is implied that the SMBH mass for \tywin{} is an order of magnitude larger than that of \bran{}. The mass ($6 \, M_\odot$) and radius ($R_\ast \simeq 4 R_\odot$) of the disrupted star were assumed to be much larger as well. This is natural considering the overall larger energy budget and longer duration of the flare, as long as the available energy is proportional to a fixed fraction of the star's mass. This assumption is also consistent with the prediction that the contribution of super-solar stars to the TDE rate increases with $M$ \cite{Kochanek:2016zzg}.

Assuming that the Schwarzschild time $\tau_S \simeq 630 \, \mathrm{s} \, (M_{\mathrm{SMBH}}/(10^7 \, M_\odot))$ translates into the time variability of the engine, we assumed a variability timescale $t_v \simeq 1000 \, \mathrm{s}$. This means that internal shocks from the intermittent engine will form at a distance (collision radius) $R_C \simeq 2 \, \Gamma^2 \, t_v\,c \simeq 10^{16} \, \mathrm{cm}$ for our choice of $\Gamma=14$, which is roughly comparable to the blackbody (BB) radius at peak times, see Table~\ref{tab:double_bb}. Since $t_v$ is proportional to the SMBH mass, a larger value of $R_C$ was expected for \tywin{} compared to \bran{} -- consistent with the hypothesis of a larger system. Similar to \cite{Winter:2020ptf}, it was postulated that the production and the isotropization radius roughly evolve like the BB radius to enhance the late-time neutrino production efficiency, which means that our model follows the BB fit values from Table~\ref{tab:double_bb} (linearly interpolating, and extrapolating to constant values outside the observation epochs).

\begin{figure}
    \centering
    \includegraphics[width=\columnwidth]{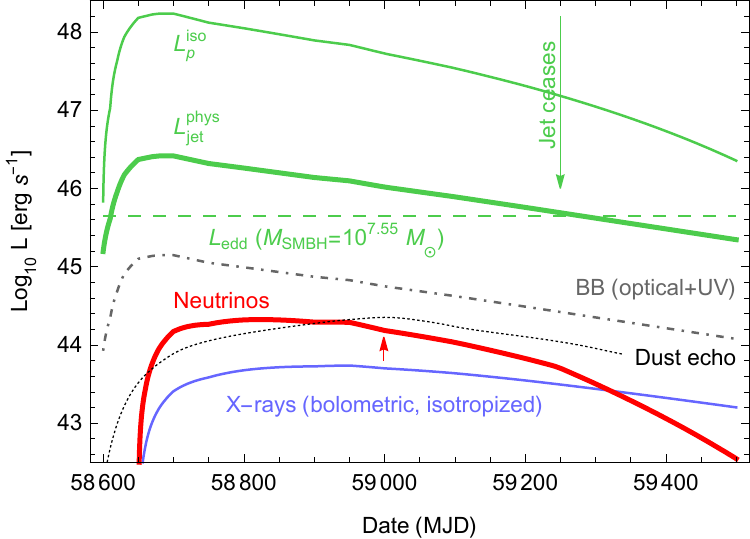}
    \caption{Evolution of luminosities as a function of MJD. The BB curve (black-dashed) extrapolates the observed lightcurves, normalized to the O+UV luminosity at peak (see Table~\ref{tab:double_bb}). The Eddington luminosity is shown as well (green-dashed), the physical jet luminosity (thick-green) is assumed to follow the BB luminosity, peaking at $6\, L_{\mathrm{edd}}$. The isotropic-equivalent non-thermal proton luminosity $L_p^{\mathrm{iso}} \simeq (2 \, \Gamma^2) \, \varepsilon\, L^{\mathrm{phys}}_{\mathrm{jet}}$, where $(2 \, \Gamma^2)$ is the beaming factor and $\varepsilon \simeq 0.2$ is the transfer (dissipation) efficiency from jet kinetic energy into non-thermal radiation (here assumed to be dominated by baryons). The X-ray luminosity (purple) is the bolometric isotropized luminosity available for neutrino production. The neutrino lightcurve (red) is a result of the computation. We have marked the actual neutrino arrival time with a red arrow and show the dust echo curve from Fig.~1 (main text) for comparison.}
    \label{fig:lumis_dbb}
\end{figure}

The jet luminosity was assumed to be $L^{\mathrm{phys}}_{\mathrm{jet}} \simeq 6\, L_{\mathrm{edd}}$ at the peak (slightly lower than in \citep{Dai:2018jbr} to avoid too extreme mass estimates for the disrupted star), and evolving like the BB luminosity (see fig. \ref{fig:lumis_dbb}); the resulting total energy into the jet was $\simeq 0.5 \, M_\odot$, which is roughly consistent with the assumed $6 \, M_\odot$ of the disrupted star.

The isotropic-equivalent non-thermal proton luminosity (thin green curve in Fig.~\ref{fig:lumis_dbb}) $L_p^{\mathrm{iso}} \simeq (2 \, \Gamma^2) \, \varepsilon\, L^{\mathrm{phys}}_{\mathrm{jet}}$, where $(2 \, \Gamma^2)$ is the beaming factor and $\varepsilon \simeq 0.2$ is the transfer (dissipation) efficiency from jet kinetic energy into non-thermal radiation (here assumed to be dominated by baryons). The jet ceases once $L^{\mathrm{phys}}_{\mathrm{jet}} \lesssim L_{\mathrm{edd}}$ (see e.g., \cite{Rees:1988bf,2012ApJ...760..103D}). Protons were assumed to be accelerated in internal shocks to an $E^{-2}$ spectrum with the maximal energy being obtained from comparing the acceleration rate with energy losses and escape rates. The resulting number of expected muon neutrino events is 0.027.

It is an interesting feature of the model that back-scattered (isotropized) X-rays stemming from the disk serve as external target photons. In this scenario, travel times in the system can lead to a delay of the neutrino signal from building up the external radiation field. Compared to \bran{}, there are several possibilities which may induce this isotropization: (sublimated) dust, an outflow, or even the broad line region. For example, if the X-ray scattering occurs near the radius of the IR emission, the delay due to the light travel (photons traveling twice the distance, to and from the IR region) is roughly $2 \, \Delta t_c \simeq 386$ days (see the section on the dust echo for details). A drawback is a large dilution factor of the external radiation field due to the size of the isotropization region. Another possible idea is that the X-rays could be efficiently back-scattered in the inner part of the durst torus by the gas produced by dust sublimation, or some fine-tuned geometry may be at play. For example, the ramp-up timescale during which the dust echo sets in points towards a smaller region, comparable to $R_C$, which would imply only moderate dilution -- but which would come at the expense of no delayed onset of the neutrino flux. On the other hand, an outflow with $v\simeq 0.01 c$ would have a similar effect satisfying both requirements (small dilution and delayed onset), similarly to what was proposed for \bran{} \cite{Winter:2020ptf}.

Since the TDE occured within a pre-existing Seyfert galaxy, it is also possible that the X-ray scattering occured in the broad line region similar to the suggestion by \citet{Murase:2014foa}. In the absence of further constraints, it was assumed that the external target photons build up over a timescale of $2 \, \Delta t_c$ after MJD 58650 (start of the broad peak) at an isotropization radius comparable to $R_C$, and that the back-scattering efficiency is 0.3, which is closest to the outflow scenario. Note that if the obscuration were at play at early times (where strong X-ray limits exist, see Fig.~\ref{fig:complete_lightcurve}) and the late-term signal was indicative for the flux, then the unobscured X-ray flux could be much higher -- which would allow for a larger dilution factor.

Our results are summarized in Fig.~\ref{fig:lumis_dbb} here (luminosity evolution) and Fig.~2 in the main text (neutrino fluence as a function of energy). The neutrino lightcurve predicts an average delay similar to  the actual delay, although the predicted variation is large. The buildup of the neutrino lightcurve is delayed with respect to the BB because of the size of the system delaying the buildup of the external target photons. On the other hand, the late-term decrease of the production/isotropization radius sightly boosts the neutrino production efficiency. It is intriguing that the shape of the isotropized X-ray curve is very similar to the dust echo curve in Fig.~2 (main text); suggesting that the dust might play a role in the X-ray isotropization.

In comparison to \bran{}, the main differences in the model are the larger size of the system, the larger SMBH mass and the smaller fraction of energy going into the jet, the larger mass of the disrupted star, and a higher assumed back-scattering efficiency, because the system is expected to be covered with components of the pre-existing AGN. Furthermore, a slightly higher $\Gamma$ (than \bran{}) was assumed and a viewing angle $\theta_v=1/\Gamma$, such that Doppler and Gamma factors are identical (for \tywin{}). The neutrino event rates are relatively similar, and so are the spectra. Note that the origin of the X-ray isotropization might be different: for \bran{}, a mildly relativistic outflow -- originally associated with the radio signal -- was assumed to be the reason, since no pre-existing AGN components were expected. In the case of \tywin{}, the scattering could also come from a similar outflow, but there is no observational evidence for that and there are the other possibilities discussed earlier. In all cases, the observed time delay of the neutrino seems to scale with the size of the system.

It has been established that relativistic jets lead to radio afterglow emission. In the case of \bran{}, the radio data are best explained by sub-relativistic winds in the standard leptonic emission scenario rather than a relativistic jet~\cite{mohan_21,matsumoto_21}. The origin of the radio emission from \tywin{} is uncertain, and there is no evidence for a jet. In this regard, \cite{murase_bran} examined a hidden jet model by requiring that the relativistic jet is choked such that afterglow emission does not emerge. We do not consider this model here because of insufficient energetics to describe the neutrino event.

\bibliographystyle{apsrev4-2}

\end{document}